\documentclass[10pt,a4paper,twocolumn]{IEEEtran}
\normalsize
\usepackage{diagbox}
\usepackage[english]{babel}
\usepackage{latexsym}
\usepackage{varioref}
\usepackage{mathrsfs}
\usepackage{array}
\usepackage{hhline}
\usepackage{dcolumn}
\usepackage{tabularx}
\usepackage{algorithm}
\usepackage{algorithmic}
\usepackage{calc}
\usepackage{epsfig}
\usepackage[usenames]{color}
\usepackage{amsmath,amssymb,bigdelim}
\usepackage{t1enc}
\usepackage{pstricks,pst-node,psfrag,pst-plot}
\usepackage{pstool} 
\usepackage{graphicx,parskip,amsbsy}
\usepackage{boxedminipage}
\usepackage{indentfirst}
\usepackage{booktabs}
\usepackage[noadjust]{cite}
\usepackage{caption}
\usepackage{multirow}
\usepackage{subfigure}
\usepackage{stackengine}
\usepackage[usenames]{color}
\usepackage{clrscode}
\usepackage{url}
\usepackage{setspace}
\usepackage{fancybox}
\usepackage{ulem}
\makeatletter
\def\url@leostyle{%
  \@ifundefined{selectfont}{\def\UrlFont{\sf}}{\def\UrlFont{\small\ttfamily}}}
\makeatother
\newcommand{\tabincell}[2]{\begin{tabular}{@{}#1@{}}#2\end{tabular}}
\newcommand{\rem}[1]{}

\newcommand{\addnew}[1]{{\color{black} #1}}
\newcommand{\redfbox}[1]{{\color{black}{#1}}}
\newcommand{\rmold}[1]{}

\renewcommand{\baselinestretch}{0.99}
\providecommand{\texlivekeywords}[1]{\textbf{\textit{Index terms---}} \textbf{\small #1}}
\providecommand{\fref}[1]{Fig.\,\ref{#1}}
\providecommand{\frefp}[1]{Figs.\,\ref{#1}}
\providecommand{\sref}[1]{Sect.\,\ref{#1}}
\providecommand{\aref}[1]{Algorithm\,\ref{#1}}
\providecommand{\tref}[1]{Table\,\ref{#1}}

\def\BibTeX{{\rm B\kern-.05em{\sc i\kern-.025em b}\kern-.08em
    T\kern-.1667em\lower.7ex\hbox{E}\kern-.125emX}}

\usepackage[shortcuts,acronym]{glossaries}
\newacronym{5G}{5G}{fifth-generation}
\newacronym{MIMO}{MIMO}{multiple-input-multiple-output}
\newacronym{SISO}{SISO}{single-input-single-output}
\newacronym{BS}{BS}{base station}
\newacronym{UE}{UE}{user equipment}
\newacronym{PAS}{PAS}{power angular spectrum}
\newacronym{SMPAC}{SMPAC}{sectored MPAC}
\newacronym{SMPAC in abstract}{SMPAC}{sectored multiprobe anechoic chamber}
\newacronym{MPAC}{MPAC}{multiprobe anechoic chamber}
\newacronym{RTS}{RTS}{radiated two stage}
\newacronym{RC}{RC}{reverberation chamber}
\newacronym{HPBW}{HPBW}{half-power-beamwidth}
\newacronym{UCA}{UCA}{uniform circular array}

\newacronym{RF}{RF}{radio frequency}
\newacronym{OTA}{OTA}{over-the-air}
\newacronym{SNR}{SNR}{signal-to-noise-ratio}
\newacronym{mmWave}{mmWave}{millimeter-Wave}
\newacronym{RRM}{RRM}{radio resource management}
\newacronym{PWS}{PWS}{plane wave synthesis}
\newacronym{PFS}{PFS}{prefaded signal synthesis}
\newacronym{CSI}{CSI}{channel state information}
\newacronym{CDL}{CDL}{clustered-delay-line}
\newacronym{DUT}{DUT}{device under test}

\newacronym{CST}{CST}{computer simulation technology}

\newacronym{VNA}{VNA}{vector network analyzer}
\newacronym{CTF}{CTF}{channel transfer function}
\newacronym{MPC}{MPC}{multipath component}
\newacronym{LoS}{LoS}{line-of-sight}
\newacronym{NLoS}{NLoS}{non-line-of-sight}
\newacronym{3GPP}{3GPP}{3rd Generation Partnership Project}

\newacronym{PXI}{PXI}{PCI eXtensions for Instrumentation}
\newacronym{mMIMO}{mMIMO}{massive multiple-input-multiple-output}
\newacronym{UAV}{UAV}{unmanned aerial vehicle}
\newacronym{SDR}{SDR}{Software Defined Radio}
\newacronym{CIR}{CIR}{channel impulse response}
\newacronym{SINR}{SINR}{signal-to-interference-plus-noise ratio}
\newacronym{USRP}{USRP}{Universal Software Radio Peripheral }
\newacronym{APDP}{APDP}{averaged power delay profile}
\newacronym{PDP}{PDP}{power delay profile}
\newacronym{SAGE}{SAGE}{space-alternating generalized expectation-maximization}
\newacronym{HRPE}{HRPE}{high-resolution-parameter-estimation}
\newacronym{LTE}{LTE}{Long Term Evolution}
\newacronym{FDD}{FDD}{frequency division duplex}
\newacronym{EM}{EM}{expectation-maximization}
\newacronym{AIC}{AIC}{Akaike information criterion}
\newacronym{KPM}{KPM}{K-Power-Means}
\newacronym{MCD}{MCD}{multipath component distance}
\newacronym{GMM}{GMM}{Gaussian-mixture-model}
\newacronym{CH}{CH}{Cali\~nski-Harabasz}
\newacronym{DB}{DB}{Davies-Bouldin}
\newacronym{RMS}{RMS}{root-mean-square}
\newacronym{A2G}{A2G}{air-to-ground}
\newacronym{C2}{C2}{command and control}
\newacronym{AoA}{AoA}{azimuth of arrival}
\newacronym{OLPC}{OLPC}{open loop power control}
\newacronym{PLE}{PLE}{path loss exponent}
\newacronym{RB}{RB}{resource block}
\newacronym{TUE}{TUE}{terrestial-UE}
\newacronym{UUE}{UUE}{UAV-UE}
\newacronym{FDPS}{FDPS}{frequency domain packet scheduling}
\newacronym{PF}{PF}{proportional fair}
\newacronym{CDF}{CDF}{cumulative distribution function}
\newacronym{ISD}{ISD}{Inter-site distance}
\newacronym{GP}{GP}{geometrical programming}
\newacronym{SE}{SE}{spectrum efficiency}
\newacronym{SC}{SC}{single-carrier}
\newacronym{KTT}{KTT}{Karush–Kuhn–Tucker}
\newacronym{TTI}{TTI}{transmision time interval}
\newacronym{QoS}{QoS}{quality of service}
\newacronym{NOMA}{NOMA}{non-orthogonal multiple access}
\newacronym{SCA}{SCA}{successive convex approximation}

\begin{document}

	\pagestyle{plain}
  \title{Power Allocation for Uplink Communications of Massive Cellular-Connected UAVs} 

	\author{Xuesong Cai,~\IEEEmembership{Senior Member,~IEEE,} Istv\'an Z. Kov\'acs,~\IEEEmembership{Member,~IEEE,} Jeroen Wigard, Rafhael Amorim, Fredrik~Tufvesson,~\IEEEmembership{Fellow,~IEEE,} and Preben E. Mogensen

		\thanks{
      
 The final version can be found in IEEE Transactions on Vehicular Technology. Digital Object Identifier 10.1109/TVT.2023.3244653
 
      X. Cai and F. Tufvesson are with the Department of Electrical and
      Information Technology, Lund University, 22100 Lund, Sweden (email:
      xuesong.cai@eit.lth.se; fredrik.tufvesson@eit.lth.se). 
    

I. Z. Kov\'acs, J. Wigard and R. Amorim are with Nokia Standards, 9220 Aalborg, Denmark (email: istvan.kovacs@nokia.com; jeroen.wigard@nokia.com; rafhael.medeiros\_de\_amorim@nokia.com).


P. E. Mogensen is with the 
Department of Electronic Systems, Aalborg University, 9220 Aalborg, Denmark, and Nokia Standards, 9220 Aalborg, Denmark (e-mail:
pm@es.aau.dk).


This paper has been submitted to IEEE for possible publication. 

}

	}

\markboth{IEEE Transactions on Communications}%
{Submitted paper}

\maketitle \thispagestyle{plain}

\begin{abstract} 
  \addnew{Cellular-connected \acf{UAV} has attracted a surge of research interest in both academia and industry. To support aerial user equipment (UEs) in the existing cellular networks, one promising approach is to assign a portion of the system bandwidth exclusively to the \acs{UAV}-\acsp{UE}. This is especially favorable for use cases where a large number of \acs{UAV}-\acsp{UE} are exploited, e.g., for package delivery close to a warehouse. Although the nearly \acf{LoS} channels can result in higher powers received, \acsp{UAV} can in turn cause severe interference to each other in the same frequency band.
  In this contribution, we focus on the uplink communications of massive cellular-connected \acsp{UAV}. 
  Different power allocation algorithms are proposed to either maximize the minimal \acf{SE} or maximize the overall \acs{SE} to cope with the severe interference based on the successive convex approximation (SCA) principle. One of the challenges is that a \acs{UAV} can affect a large area meaning that many more \acs{UAV}-\acsp{UE} must be considered in the optimization problem, which is essentially different from that for terrestrial \acsp{UE}. The necessity of single-carrier uplink transmission further complicates the problem. Nevertheless, we find that the special property of large coherent bandwidths and coherent times of the propagation channels can be leveraged. The performances of the proposed algorithms are evaluated via extensive simulations in the full-buffer transmission mode and bursty-traffic mode. Results show that the proposed algorithms can effectively enhance the uplink \acsp{SE}. This work can be considered the first attempt to deal with the interference among massive cellular-connected \acs{UAV}-\acsp{UE} with optimized power allocations.}

\end{abstract}
\texlivekeywords{Interference, power control, geometrical programming, successive convex approximation, and UAV communications.}
\IEEEpeerreviewmaketitle

\section{Introduction}

\addnew{Due to the rapid development of \acp{UAV} in reducing their costs, sizes, weights and energy consumption, the \ac{UAV} assistance paradigm \cite{ALZAHRANI2020102706}, i.e., using \acp{UAV} to support many applications and terrestrial networks, users, and communicating entities, has attracted significant attention.} 
Specifically, deployed as aerial \acp{BS} \cite{9170560}, \acp{UAV} can quickly provide wireless connections, e.g., in the damaged areas after disasters for rescue purposes. \addnew{By leveraging the nearly \ac{LoS} radio propagation channels, \acp{UAV} can also be utilized as relays to enable energy-efficient sensing, internet-of-things and coverage improvement \cite{9121255,9115248,9566766}}. Moreover, other applications such as network optimization, smart agriculture, forest monitoring, goods delivery, etc. \cite{hayat2016survey,kumar2018unmanned} are becoming autonomous and convenient with the help of \acp{UAV}. 
Meanwhile, \acp{UAV} as new aerial \acp{UE} connected to the existing cellular networks, i.e., cellular-connected \acp{UAV} \cite{8470897,8918497}, have been identified as an important paradigm shift. The current almost everywhere cellular networks are expected to provide reliable \ac{C2} and high-throughput payload communications for \ac{UAV}-\acp{UE} to enable, e.g., beyond visual \ac{LoS} flight tasks and high-definition video streaming. 
However, although the nearly \ac{LoS} channels are advantageous to receive higher powers, \acp{UAV} can in turn cause severe interference. 
This becomes more critical with the rapid growth of the number of \acp{UAV} that are connected to cellular networks, significantly affecting the connectivity of both aerial and terrestrial \acp{UE}.

\addnew{
Different techniques have been investigated to provide reasonable services to \acp{UAV} and terrestrial \acp{UE}. In \cite{8528463,8869706,8470897}, the authors proposed to exploit massive \ac{MIMO} available at \ac{BS} side to point beams towards intended \acp{UE} and more sophistically place spatial nulls to \acp{UE} in other cells that are vulnerable to the interference. In such a way, beamforming gain and spatial multiplexing gains can be harvested to improve wireless connectivity. Although massive \ac{MIMO} at \ac{BS} is promising, it requires large investments and takes time to upgrade the infrastructure. As a cheaper and fast alternative, implementing a beamforming system or directional antennas on board the \ac{UAV} was proposed in \cite{8301389,9082692}. The authors have shown its effectiveness to increase the uplink performance of \ac{UAV} communications. In \cite{9777746}, the authors proposed to exploit the passive beamforming of intelligent reflecting surfaces (IRSs) to achieve an optimal performance tradeoff between UAV-UEs and terrestrial UEs. In \cite{9925817}, a deep Q-learning framework is utilized to tune the standard fixed power allocation algorithm for the UAV to mitigate interference, which requires the UAV to only update its coordinates and received \ac{SINR} continuously. In \cite{9915685}, the effectiveness of three different interference mitigation methods including power control, beamforming and coordinated multi-point transmission was compared. In \cite{9878252}, the authors considered the case of one \ac{BS} serving a cluster of \acp{UAV}, and the sum rate or the minimum rate of \acp{UAV} was optimized by power control. The authors in \cite{9427545} investigated the uplink SINR distribution of UAV-UEs and terrestrial UEs based on the technique of stochastic geometry. It is found that there exists an optimal height of UAVs for SINR enhancement. In \cite{8811738}, by jointly optimizing the \ac{UAV}'s uplink cell associations and power allocations, the weighted sum rate of the \ac{UAV} and terrestrial \acp{UE} was optimized. In \cite{9099899}, idle \acp{BS} without serving any other \acp{UE} in the \ac{UAV}'s communication channel were exploited to help co-channel \acp{BS} mitigate the interference. In \cite{8763928}, a cooperative interference cancellation strategy that exploits the existing backhaul links among \acp{BS} was proposed for sum-rate maximization. Specifically, a  multi-antenna UAV sends multiple data streams to its serving \acp{BS}, which are then forwarded to the backhaul-connected \acp{BS} that serve terrestrial UEs for interference cancellation. In \cite{9324909}, the authors provided a good summary of different interference coordination and mitigation algorithms for the coexistence of UAV-UEs and terrestrial UEs. A comparison of the above works is shown in Table\,\ref{TableQ}. } 

{ 
\begin{table*}
\centering
\caption{\addnew{Investigations of interference mitigation for cellular-connected \acp{UAV}.}\label{TableQ}}
  \scalebox{0.81}{
    \redfbox{
\begin{tabular}{cp{8cm}p{5cm}p{5cm}p{6cm}}
\hline\hline 
 References &  \tabincell{c}{Techniques}  &  \tabincell{c}{Methodology}   &   UE types  \\ \hline 
  \cite{8528463,8869706,8470897}  & Massive MIMO, beamforming, spatial nulling   &   Simulations  & UAV-UEs and terrestrial UEs   \\
\cite{8301389,9082692}  &    Directional antennas onboard a UAV  &   Measurement verification   &   One UAV and terrestrial UEs       \\
\cite{9777746}    &   IRSs, passive beamforming    &   Sum-rate optimization   &    One UAV and terrestrial UEs  \\
\cite{9925817}  &    Power control            &    Deep Q-learning   &   UAVs     \\ 
\cite{9915685,9324909} &    Power control, beamforming, coordinated multipoint, frequency reuse, interference cancellation, etc.    &    Simulations  & Several UAV-UEs and terrestrial UEs   \\
\cite{8811738}  &   Uplink cell association, power control    &    Weighted sum-rate optimization   &     One UAV and terrestrial UEs  \\  

  \cite{8763928,9099899}     &   Backhaul, multi-beam UAV, cooperative interference cancellation    &   Weighted sum-rate optimization  &   One UAV and terrestrial UEs      \\
\hline
\end{tabular}}
}
\end{table*}
}

Although the above-mentioned approaches have demonstrated potential in improving the performance of serving flying \acp{UAV} and protecting terrestrial \acp{UE}, \textit{{they do not well consider the case where the density of \acp{UAV} is very high,}} which is probable in the near future. The basic assumption in these investigations, e.g. \cite{9324909,9099899,8811738,9082692}, is that only a few (even only one) \acp{UAV} exist in a large area. However, as predicted by Federal Aviation Administration (FAA), the number of commercial \ac{UAV} fleets can reach up to 1.6 million by 2024 \cite{6guav,FAA}. In \cite{aerialporfilev1}, a typical use case of cellular-connected \acp{UAV} is recognized as the package delivery, e.g., the Amazon Prime Air project in the early stage, with a medium to high \ac{UAV} density. It is estimated that several \acp{UAV} can exist per square mile near to the warehouse or operations center around 2022, which means that each cell (in the hotspot areas) can have several active \ac{UAV}-\acp{UE}. Developing robust interference mitigation techniques for \textit{massive} \ac{UAV}-\acp{UE} deployment scenarios is also considered a key open problem in \cite{8660516}. \textit{{Therefore, identified as an important communication scenario that has seldomly been addressed in the literature, we in this work focus on the uplink communication for \ac{UAV}-\acp{UE} of high densities, i.e., a massive number of \ac{UAV}-\acp{UE}.}} 

\begin{figure}
  \begin{center}
  \includegraphics[width=0.45\textwidth]{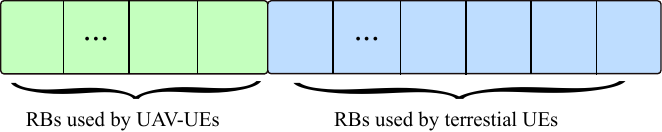}
  \end{center}
  \caption{UAV-UEs and terrestrial UEs use orthogonal RBs.\label{fig:bandallo}}
  \end{figure}

Due to the nearly-\ac{LoS} channels, these \ac{UAV}-\acp{UE} can cause severe interference to terrestrial \acp{UE} in a rather large area. As demonstrated in \cite{8660516}, the connectivity probability of terrestrial \acp{UE} can be lowered significantly, e.g. from 0.95 to 0.7 with the number of \acp{UAV} increasing. To avoid the many \acp{UAV} negatively affecting the performance of terrestrial \acp{UE}, we assume, similarly as discussed in \cite{8758988,8869706}, that \ac{UAV}-\acp{UE} and terrestrial \acp{UE} use orthogonal \acp{RB} in each transmission interval as illustrated in \fref{fig:bandallo}.  This spectrum sharing has been suggested in \cite{9115898} as the most suitable strategy for maintaining a guaranteed rate for \acp{UAV} and high performance of terrestrial \acp{UE} if the number of \acp{UAV} is large. The investigation in \cite{8869706} also shows its effectiveness. In \cite{9681624}, the authors also suggest the necessity in the future of dedicated \ac{A2G} cellular networks without serving terrestrial \acp{UE}. Therefore, the crux we aim to solve in this work is to optimize the uplink performance of a large number of \acp{UAV}, coping with the inter-cell interference among them. It is worth noting that although the problem seems similar to the conventional case where all \acp{UE} are terrestrial and \addnew{different algorithms \cite{4275017,MAPEL,RR,ADP,BPC,DNN} are available in the literature with local optimal solutions}, \textit{it does differ mainly in two aspects. \textit{\textbf{i)}} The \ac{A2G} channels between the \ac{UAV}-\acp{UE} and \acp{BS} are nearly \ac{LoS}. This bring new properties and performance behavior of \acp{UAV}; \textit{\textbf{ii)}} Many more, e.g., tens or even up to hundreds of cells need to be considered compared to only a few cells in the terrestrial case, which can lead to exponentially increasing complexity.} To this end, {the main contributions and novelties of this paper are summarized as follows:}
\begin{itemize}

  \item To deal with the exponential complexity caused by the massive number of \acp{UAV} in the optimization problem, a novel \ac{SCA} technique is applied based on the principle of \ac{GP}.

  \item By exploiting the special characteristics of the nearly \ac{LoS} \ac{UAV} propagation channels, different algorithms applied in the frequency domain and/or the time domain are proposed to optimize the \ac{SE} of the system. The important and practical constraint, i.e., uplink \ac{SC} constraint\footnote{Orthogonal Frequency Division Multiple Access (OFDMA) has been a standard radio access technique in 4G LTE and 5G networks. However, a major issue of OFDMA is a high peak-to-average power ratio (PAPR), which is unfriendly to mobile UEs whose power consumption is a key consideration. Thus, \ac{SC}-OFDMA with significantly lower PAPR has been an alternative for mobile UEs for uplink transmission. Because the waveform is essentially single-carrier, the occupied bandwidth has to be continuous and with the same power level, which is the \ac{SC} constraint considered herein.}, is also considered.

\item  Extensive simulations are performed to evaluate the performances of the proposed algorithms. Both full-buffer transmission and bursty-traffic modes are considered. Moreover, clustered application of the algorithms is also discussed in compromising between performance and signaling overhead. The numerical results show the effectiveness of these algorithms and provide important insights into the practical system design.


\end{itemize}



  The rest of this paper is structured as follows. In \sref{sect2}, low altitude \ac{A2G} channel properties and preliminary understanding of the uplink \ac{UAV} communications are discussed. In \sref{sect3}, different power allocation algorithms are proposed and elaborated. \sref{sect4} presents the extensive simulations in full-buffer transmission mode and bursty-traffic mode for evaluating the proposed algorithms. Detailed discussions are also included. Finally, conclusive remarks are given in \sref{sect5}.

\section{The uplink communications of massive cellular-connected UAVs} \label{sect2}

In this section, we discuss the characteristics of the propagation channels among \acp{UAV} and terrestrial \acp{BS}, and a preliminary analysis of power allocation, interference and scheduling is also included.


\subsection{Low altitude {A2G} propagation channels\label{sect:channels}}

Many measurement-based investigations such as \cite{3gppenhance,7936620,8807190,9170768,xuesongjsac} have demonstrated that the \ac{A2G} propagation channels at higher heights are nearly \ac{LoS}. For example, 3GPP \cite{3gppenhance} suggests a \ac{LoS} probability of 1 for rural areas and moderately high \ac{UAV} altitude larger than 40\,m, which is also verified by the measurements in \cite{xuesongjsac} for even lower heights. In \cite{7936620}, \acp{PLE} were found to be almost free-space values as 2.1 and 2 at 60\,m and 120\,m, respectively. In \cite{8807190,9170768}, the Rician K-factors of the \ac{A2G} channels were found to be large for most cases at higher heights with a mean value around 15\,dB. Although in few cases, the K-factor could be smaller due to, e.g., the reflections from buildings \cite{9170768,xuesongjsac} in urban or industrial areas, spatial analysis in \cite{xuesongjsac} for the \ac{A2G} channels has shown that the additional cluster(s) are usually separable to the \ac{LoS} cluster in the azimuth domain with angle differences larger than 60$^\circ$. This means that if a \ac{UAV} is equipped with directional antennas, even the channel with several clusters can probably be simplified as a single-cluster channel with a high K-factor. This is a reasonable expectation since as shown in \fref{fig:sinrs} in \sref{sect:intferences}, using omnidirectional antennas onboard \acp{UAV} can cause rather severe interference.
The authors in \cite{9082692} exploited an array consisting of six directional antennas each with a \ac{HPBW} of 60$^\circ$ to cover the whole 360$^\circ$ azimuth, and the best antenna was triggered for the communication. This relatively simple and low-cost switching strategy of directional antennas was able to significantly improve the performance. 
\addnew{Given the above reasoning, it can be inferred that the nearly \ac{LoS} or single-cluster \ac{A2G} channels are quite flat in a certain bandwidth. This has also been demonstrated by the ultra-wideband \ac{A2G} channel measurements \cite{7842372} where the coherent bandwidths of the channels are found to be at least 100\,MHz at different scenarios and heights.} Moreover, for the same reasons and the relatively low speeds of low-altitude \acp{UAV}, the \ac{A2G} channels have relatively large coherent time, e.g. tens of ms or even longer \cite{ZeyuTWC}. \addnew{For example, for 20 LTE symbols that last less than 2\,ms, a UAV at a speed of 10\,m/s moves less than 2\,cm, i.e, only 1/12 wavelength at 1.8\,GHz. It is reasonable to assume unchanged channel gains during the period.}

\subsection{Inter-cell interference\label{sect:intferences}}

\begin{figure}
  \begin{center}
    \redfbox{\subfigure[]{\includegraphics[width=0.23\textwidth]{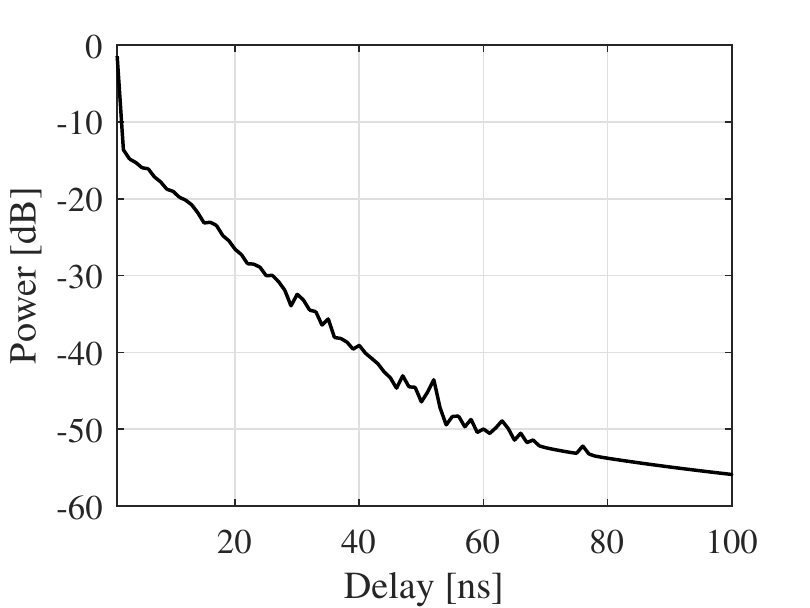}{\psfrag{F(x)}[l][l][0.65]{Omnidirectional}}}}
    \redfbox{\subfigure[]{\includegraphics[width=0.23\textwidth]{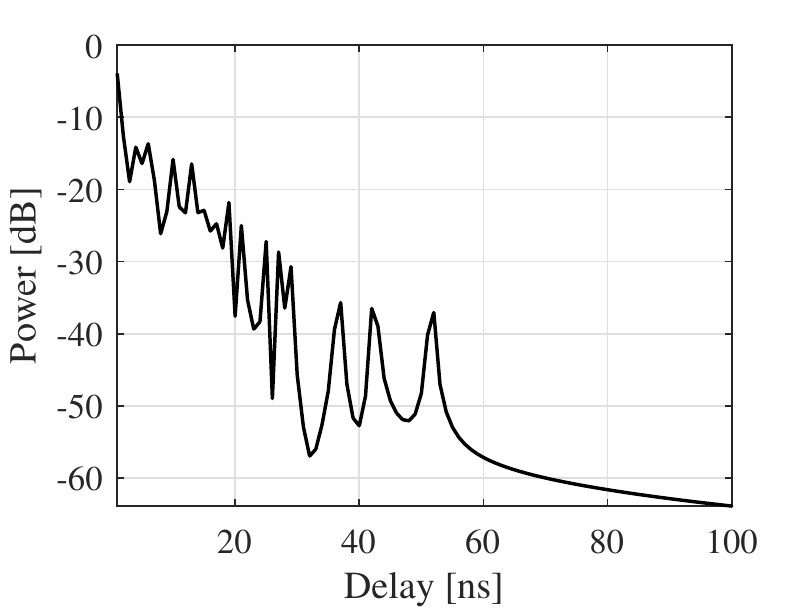}}}
    \redfbox{\subfigure[]{\includegraphics[width=0.23\textwidth]{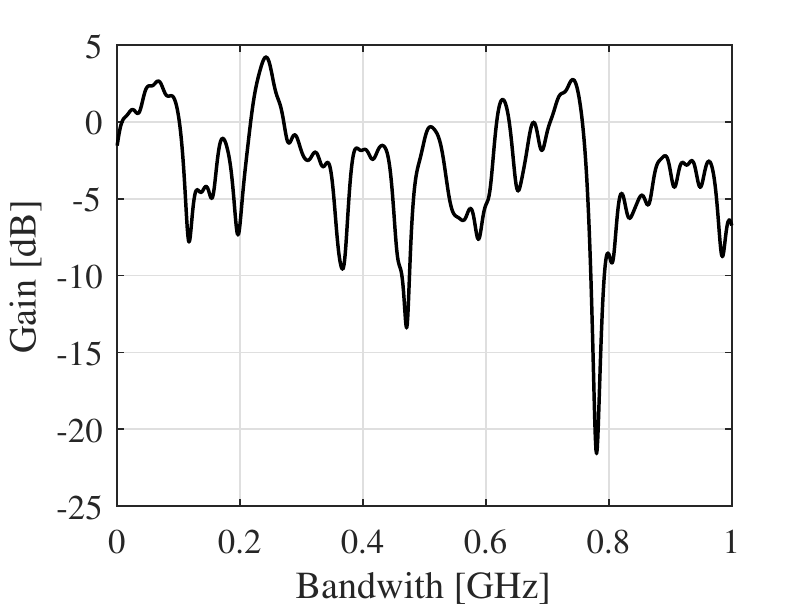}}}
    \redfbox{\subfigure[]{\includegraphics[width=0.23\textwidth]{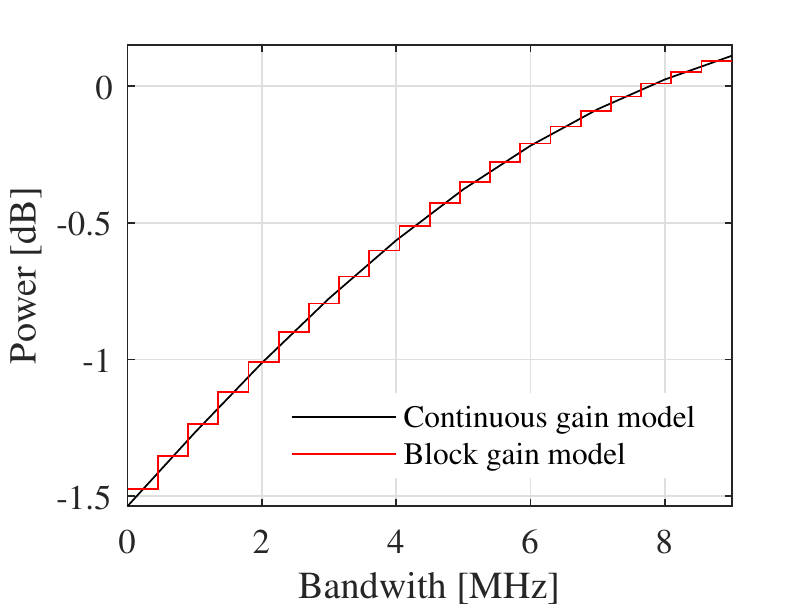}}}
  \end{center}
  \caption{{Tapped-delay-line channel model. (a) Power delay profile. (b) A realization of the channel.  (c) Channel gains across a bandwidth of 1\,GHz for one of the channel realizations as shown in (b). (d) Channel gains across a bandwidth of 8\,MHz for one of the channel realizations as shown in (b). \label{fig:channelmodels}}}
  \end{figure}

\begin{figure}
    \begin{center}
    \includegraphics[width=0.45\textwidth]{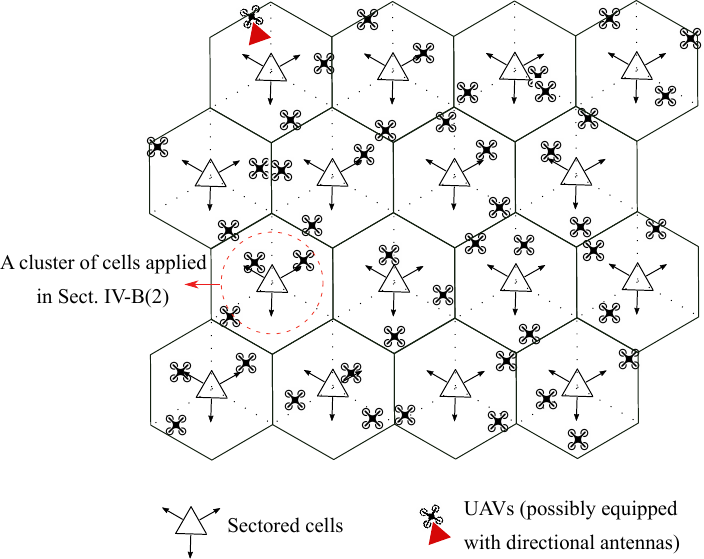}
    \end{center}
    \caption{An example realization of 48 UAVs located in the 48 sectored cells. 
    \label{fig:cells}}  
    \end{figure}

We in this subsection show the inter-cell interference among \ac{UAV}-\acp{UE} via a simulation. The {simulation parameters} are included in \tref{tab:sim_parameter}. \addnew{Briefly, we consider a rural scenario where 48 sectored cells, as illustrated in \fref{fig:cells}, were deployed.} The \ac{BS} heights were 35\,m, the downtilts of the sector antennas were 8.5$^\circ$, and the \ac{ISD} was 2\,km. In each realization, 48 \ac{UAV}-\acp{UE} were randomly generated in the 48 cells at the height of 60\,m. \addnew{For the large-scale fading, i.e., path loss and shadowing, we reproduce the measurement results obtained in \cite{7936620}. Specifically, the PLE, standard deviation of the shadowing [dB] and the intercept point at 1\,m are set as 2.1, 4.4\,dB and 32.8\,dB for the height of 60\,m, respectively. For the characteristics of multipath components that account for small-scale fading, we reproduce channels that were observed in the ultra-wideband \ac{A2G} channel measurements in \cite{7842372}. Specifically, tapped-delay-line channels are statistically realized according to the procedure as specified in \cite{7489014,3GPP38901}. The generated channels are illustrated in Fig.\,\ref{fig:channelmodels}, where the power delay profile and the channel gain evolving across the frequency band are intended to be similar to that in \cite{7842372}. Moreover, the considered bandwidth in this work is set as 9\,MHz, i.e., half the maximum effective bandwidth of a LTE system. It can be observed from Fig.\,\ref{fig:channelmodels}(d) that the channel gain is quite flat. Note that the discretized/blocked channel gains with 20 steps are finally applied.}

\begin{figure}
    \begin{center}
    \redfbox{\includegraphics[width=0.43\textwidth]{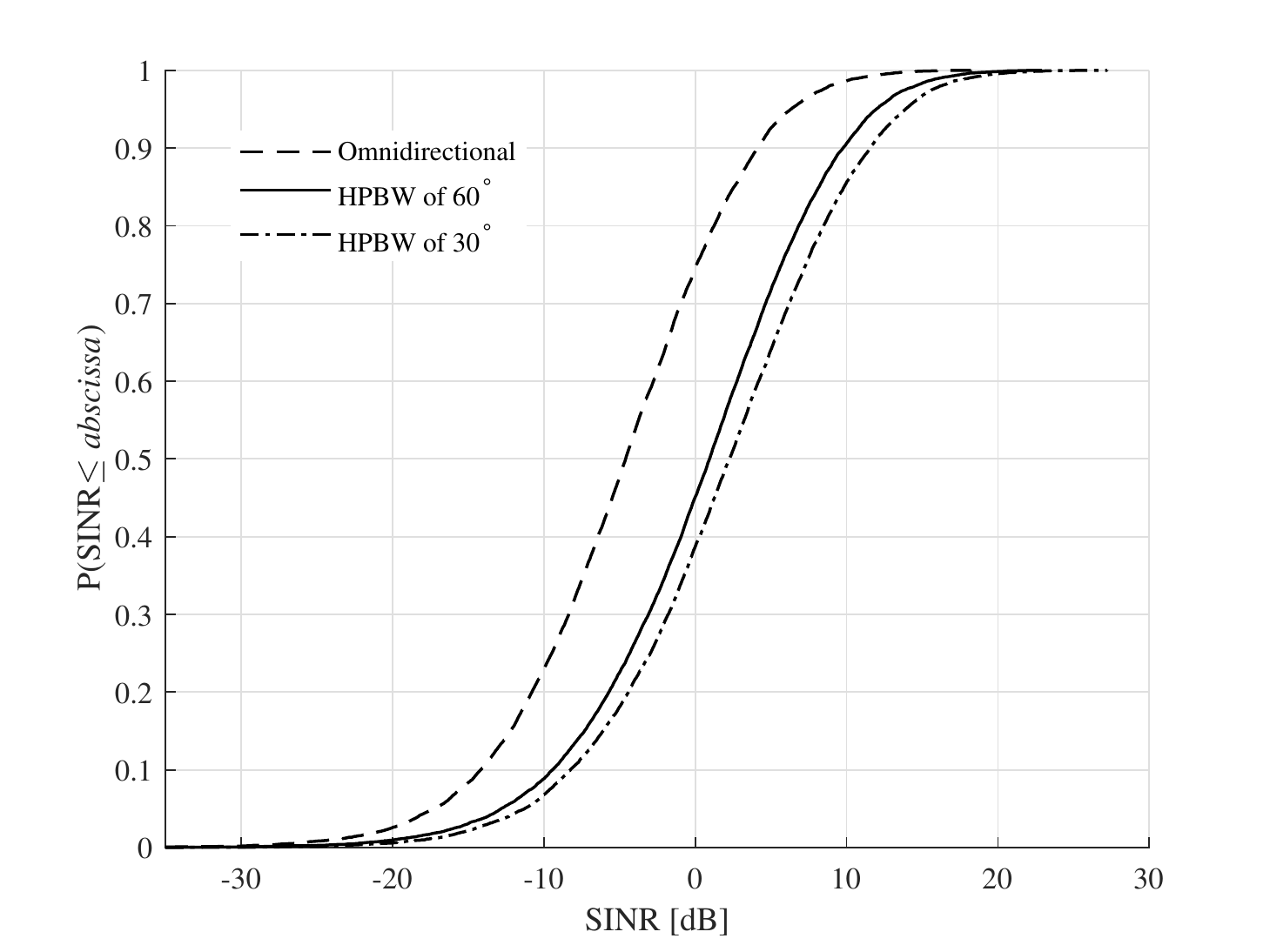}}
   \end{center}
    \caption{SINR distributions using different antennas onboard \acp{UAV}.\label{fig:sinrs}}
    \end{figure}

The power control algorithm applied in the simulation was \ac{OLPC}. Specifically, the transmitted power $P_\text{u}$ in dBm is obtained as \cite{maggi2020bayesian}
\begin{equation}
  \begin{aligned}
  P_\text{u}= \min\{P_\text{max}, P_0 + 10\log_{10}(M_{\text{RB}}) + \alpha PL\}
  \end{aligned} 
  \label{eq:olpc}
  \end{equation} where $P_\text{max}$ is the configured maximum output power (which was 23\,dBm\footnote{\addnew{According to the standard 3GPP TS 38.101-1 \cite{3GPPTS38101}, the maximum transmission power of UEs operating at sub-6\,GHz bands is 23\,dBm. Currently, there are no standards dedicated to UAV-UEs. Therefore we use 23\,dBm in this work.}}), $PL$ is path loss, $M_{\text{RB}}$ is the number of \acp{RB} allocated for this \ac{UE}, $\alpha$ is the fractional power control compensated parameter, and $P_0$ is the power received at one \ac{RB} if path loss is fully compensated. {Note that $P_0$ and $\alpha$ were optimized by exhaustive searching \addnew{(within [-90, -70]\,dBm and [0.5, 1] respectively)} in the simulation scenario as illustrated in \fref{fig:cells}.} Five hundred realizations were performed to obtain a distribution of the \acp{SINR} of individual \acp{UAV}. \fref{fig:sinrs} illustrates the \acp{CDF} of \acp{SINR} with \acp{UAV} equipped with omni-directional antennas, direction antennas of 60$^\circ$ \ac{HPBW} and directional antennas of 30$^\circ$ \ac{HPBW}, respectively. It can be observed from \fref{fig:sinrs} that omni-directional antennas onboard \acp{UAV} can usually cause severe interference. Using directional antennas with \acp{HPBW} of 60$^\circ$ can improve the \acp{SINR} significantly, due to the directional radiation pattern suppressing interference.  However, the improvement from 60$^\circ$-\ac{HPBW} antennas to 30$^\circ$-\ac{HPBW} antennas is much less than that from omni-directional to 60$^\circ$-\ac{HPBW} antennas. This means that using 60$^\circ$-\ac{HPBW} antennas could be a suitable choice considering the performance improvement and complexing increase. Nevertheless, interference can still be sometimes high for some \ac{UAV}-\acp{UE}. 

  \begin{table}
    \centering
    \caption{Important parameters configured in the simulation.}
    \scalebox{1}{
    \begin{tabular}{ll}
    \hline
    \hline
    \multicolumn{2}{c}{{{{\textit{Main parameters applied in the simulation}}}}}\\
    \hline
    Network scale  & 48\,cells      \\
    Cell type & Sectored hexagon \\
    \ac{ISD} & 2\,km \\
    BS height & 35\,m \\
    Sector antenna \acp{HPBW} & 120$^\circ$ in azi.; 13$^\circ$ in ele. \\
    Sector antenna downtilt & 8.5$^\circ$ \\
    UAV height & 60\,m \\
    Maximum uplink power & 23\,dBm\\
    Noise power density & -174\,dBm/Hz\\
    $P_0$ & -81\,dBm\\
    $\alpha$ & 0.9\\
    \addnew{PLE} & \addnew{2.1}\\
    \addnew{Path loss at 1\,m} & \addnew{32.8\,dB}\\
    \addnew{Std. of shadowing}  & \addnew{4.4\,dB}\\
    \addnew{Tapped-delay-line channel} &  \addnew{\cite{7842372}}\\
    \hline\hline
    \end{tabular}}
    \label{tab:sim_parameter}
    \end{table}

    \subsection{Packet scheduling for multiple \ac{UAV}-\acp{UE} in one cell\label{sect:pf}}

    We use both time-domain and frequency-domain packet scheduling for multiple \ac{UAV}-\acp{UE} in the same cell. Conventionally in cellular networks, e.g. LTE, packet scheduling is achieved via the so-called \textit{\ac{FDPS}} \cite{5671623,5062197} (note that time-domain is inherently considered in \ac{FDPS}). The instantaneous channel conditions of \acp{UE} at all \acp{RB} in a cell are the inputs of \ac{FDPS}, according to which different bandwidth portions are dynamically allocated to different \acp{UE} to better exploit the available frequency and user diversity, meanwhile fulfilling the uplink \ac{SC} constraint. A well-known and commonly used approach is the \ac{PF}-\ac{FDPS} that maximizes the long-term sum logarithmic utility of the system \cite{5062197,4786509}. Specifically, consider a cell with a set $\mathcal{K} = \{1,\cdots,K\}$ of \acp{UE} and a set $\mathcal{C} = \{1,\cdots C\}$ of \acp{RB}, the $c$th \ac{RB} at time $t$ is allocated to the \ac{UE} with index $\hat{k}_c(t)$ such that 
    \begin{equation}
    \begin{aligned}
    \hat{k}_c(t) = \arg\max_k \quad r_k^c(t)/R_k(t)
    \end{aligned} 
    \end{equation}
    where $r_k^c(t)$ indicates the data rate potentially achievable by the $k$th user on \ac{RB} $c$ at time $t$, and $R_k(t)$ is the historical average rate of the $k$th \ac{UE}. In the focused communication scenario where all \ac{UAV}-\acp{UE} are with close-to frequency-flat channels, $r_k^c$'s of the $k$th \ac{UE} tend to be similar across all the $C$ \acp{RB}.\footnote{They are not exactly the same because the interference may change at different \acp{RB}.} With \ac{SC} constraint further considered e.g. as shown in \cite{5062197}, the \ac{PF}-\ac{FDPS} tends to allocate almost all \acp{RB} to the same \ac{UAV}-\ac{UE}. In other words, much less frequency diversity gain can be harvested for \ac{UAV}-\acp{UE} compared to that of terrestrial \acp{UE} with frequency selective channels. Thus, in this work, we exploit time-domain PF to decide which \ac{UAV}-\ac{UE} or \ac{UAV}-\acp{UE} to be active for transmission in the bursty-traffic mode in later \sref{sect:bursty}. After that, power allocation is optimized. 
    

\section{Proposed power allocation algorithms\label{sec:gp}} \label{sect3}


\addnew{The main motivation is to maximize the sum SE of all UAV-UEs or maximize the minimum SE among all UAV-UEs. The first emphasizes the overall performance of the system, while the second emphasizes fairness among the UAV-UEs. A compromise between the two can be done by maximizing the sum SE of the system with individual \ac{QoS} constraints considered for UAV-UEs.  
By leveraging the special channel properties of UAV-UEs, i.e., large coherent bandwidths and coherent times, the optimization can be realized in the frequency domain and/or the time domain. An intuitive example is that two UAV-UEs can transmit on the first and second half bandwidths, respectively, to avoid severe interference between them when both of them use the whole bandwidth simultaneously. This can be similarly applied in the time domain. However, with the number of UAV-UEs becoming much larger, the complexity of the problems increases fatally. Considering the constraint of \ac{SC} uplink transmission further complicates the problems.  In the sequel, we propose different algorithms and techniques in the frequency domain and/or the time domain to solve the problems, so that the performance of the uplink \ac{UAV} communications can be enhanced. 
}




\subsection{Frequency-domain maximization of the minimum \ac{SE} of \ac{UAV}-\acp{UE}\label{sec:fdmaxmin}}

As discussed in \sref{sect:pf}, let us first consider the case where at most one \ac{UE} is scheduled simultaneously in a cell. That is, at a \ac{TTI} a set $\mathcal{N}=\{1,\cdots,N\}$ of cells are active each with one \ac{UAV}-\ac{UE} scheduled. In each cell, all the reserved bandwidth $B$ is allocated to the scheduled \ac{UAV}-\ac{UE} and divided into $\mathcal{S} = \{1,\cdots,S\}$ segments. We denote the channel gain from the \ac{UAV}-\ac{UE} in the $i$th cell to the $j$th cell \ac{BS} \addnew{at the $s$th frequency block} as $G_{ij\addnew{s}}$, and all the channel gains are contained in $\mathbf G = [G_{ijs}]$. Intuitively, $G_{ii\addnew{s}}$ is the gain of the serving link, whereas $G_{ij\addnew{s}}, i\neq j$ are that of interfering links. Note that $G_{ij\addnew{s}}$ is attributed to the path loss, shadow fading, small-scale fading, and radiation patterns of antennas. 
The power density of thermal noise is indicated by $N_0$, and the transmitted power of the $i$th \ac{UAV}-\ac{UE} on the $s$th bandwidth segment is denoted by $p_{is}$. Then the serving \ac{SINR} $\gamma_{is}$ of the $i$th \ac{UAV}-\ac{UE} on the $s$th bandwidth segment can be calculated as 
\begin{equation}
  \begin{aligned}
  \gamma_{is}(\mathbf{p}) = \frac{p_{is} G_{ii\addnew{s}}}{N_0 B S^{-1} + \sum_{j\in \mathcal{N}, j\neq i} p_{js} G_{ji\addnew{s}} }
  \end{aligned}.
  \end{equation} where $\mathbf p = [p_{is}], i\in \mathcal{N}, s\in \mathcal{S} $ is the compact notation of the power allocation for all the \ac{UAV}-\acp{UE}. The achieved \ac{SE} (bit/s/Hz) $R_i$ of the $i$th \ac{UAV}-\ac{UE} is then calculated according to Shannon formula as 
  \begin{equation}
    \begin{aligned}
    R_{i}(\mathbf{p}) = S^{-1} \sum_{s\in \mathcal{S}} R_{is} 
    \end{aligned}.
    \end{equation} 
  with
  \begin{equation}
    \begin{aligned}
    R_{is}(\mathbf{p}) = \log_{2}(1+\gamma_{is}(\mathbf p))
    \end{aligned}.
    \end{equation} The problem of maximizing the minimum \ac{SE} of all \ac{UAV}-\acp{UE} can be formulated as  
\begin{equation}
  \begin{aligned}
   \addnew{ \underset{\mathbf p}{\text{maximize}}} \quad  & \min_i R_i \quad\quad \\ 
  \text{subject to} \quad & p_{i,\text{min}} \leq p_i \leq  p_{i,\text{max}},  & \forall i \in \mathcal{N} \\
    & p_i = \sum_{s\in\mathcal{S}} p_{is},  & \forall i \in \mathcal{N}
  \end{aligned}
  \label{eq:fremaxmin}
  \end{equation} where $p_i$ is the total output power of the $i$th \ac{UAV}-\ac{UE} usually confined in a range from $p_{i,\text{min}}$ (e.g., 0) to $p_{i,\text{max}}$ (e.g., 23\,dBm), and the objective is to find the optimal $\mathbf{p}^{\ast}$ solving the problem. Note that $p_{is}$ of the $i$th \ac{UE} can be dependent on $s$. The underlying reasoning is that this allows competing \ac{UAV}-\acp{UE} transmitting on different bandwidth resources with different power densities to avoid severe interference so that the minimum \ac{SE} can be increased. An intuitive case is that two neighboring edge-\acp{UE} are allowed to transmit on the first half $B$ and the second half $B$, separately. Nevertheless, problem \eqref{eq:fremaxmin} omits the uplink \ac{SC} constraint, i.e., the same power density must be applied for continuous \acp{RB}. We discuss first how to solve \eqref{eq:fremaxmin} and consider the \ac{SC} constraint later. Problem \eqref{eq:fremaxmin} can be equivalently rewritten by introducing auxiliary variables $b_i$'s and $a_{is}$'s as
\begin{equation}
\begin{aligned}
  \addnew{\underset{\mathbf p, [a_{is}], [b_i]}{\text{maximize}}}\quad  & \min_i b_i \\ 
\text{subject to} \quad & S^{-1}\log_2 \{\prod _{s=1}^{S} (1+a_{is})\} \geq b_i , &\forall i \in \mathcal{N}\\
&  \gamma_{is} \geq  a_{is},\quad \quad\quad\quad\quad \quad\forall i \in \mathcal{N}, &\forall s \in \mathcal{S}\\
& p_{i,\text{min}} \leq p_i \leq  p_{i,\text{max}}, &\forall i \in \mathcal{N} \\
  & p_i = \sum_{s\in\mathcal{S}} p_{is}, &\forall i \in \mathcal{N}
\end{aligned}
\label{eq:fremaxmin_re}
\end{equation} which is equivalently
\begin{equation}
  \begin{aligned}
    \addnew{\underset{\mathbf p, [a_{is}], [b_i]}{\text{maximize}}}\quad  & \min_i b_i \\
  \text{subject to} \quad  & \frac{b_i} {\prod _{s=1}^{S} (1+a_{is})} \leq 1, & \forall i  \\
  & \frac{a_{is} N_0 B S^{-1} + \sum_{ j\neq i} a_{is} p_{js} G_{ji\addnew{s}} } {{p_{is} G_{ii\addnew{s}}}} \leq 1   ,&  \forall i , s\\ 
  & p_{i,\text{min}} \leq p_i \leq  p_{i,\text{max}}, & \forall i  \\
    & p_i = \sum_{s\in\mathcal{S}} p_{is}. & \forall i 
  \end{aligned}
  \label{eq:fremaxmin_2}
  \end{equation} This is a non-convex problem. If $a_{is} \gg 1$ holds for all $i$ and $s$, i.e., $\prod_{s=1}^{S} (1+a_{is})$  can be well approximated by $\prod_{s=1}^{S} a_{is}$, problem \eqref{eq:fremaxmin_2} is almost a \ac{GP}-problem where a posynomial is minimized with upper bounded posynomial and/or equality monomial constraints \cite{4275017,cai2020centralized,9301243}. However, interference in the \ac{UAV} communications are generally significant. It is also possible that a \ac{UAV}-\ac{UE} may not transmit power at some bandwidth segments to avoid interference.
  Therefore, we resort to solving the problem via a series of \acp{GP} using the single condensation principle \cite{4275017,cai2020centralized}. In our case, the term $g_i(\mathbf A) = \prod_{s=1}^{S} (1+a_{is})$ is replaced by a monomial $h_i(\mathbf A) = \prod_{s=1}^{S} c_i s_{is}^{w_{is}}$, where matrix $\mathbf A$ contains all $a_{is}$, i.e. $\mathbf A = [a_{is}]$, and $c_i$ and $w_{is}$'s are constants to be properly set. Problem \eqref{eq:fremaxmin_2} can then be solved according to \aref{al1}. It is essential that the $h_i$ meets three requirements \cite{4275017} to achieve convergence and feasibility of finally resulted power allocation, which include \textit{i)} $h_i(\mathbf A) \leq g_i(\mathbf A)$ for all $\mathbf A$. This is to guarantee the resulting power allocation always meets the original constraints. \textit{ii)} $h_i(\mathbf A_{0}) = g_i(\mathbf A_{0})$ at the condensation point $\mathbf A_{0}$. This guarantees the monotonicity of optimal values obtained in successive iterations. \textit{iii)} $\nabla {h_i(\mathbf A_{0})}= \nabla g_i(\mathbf A_{0})$ at the condensation point. This is to make sure that after convergence, \ac{KTT} conditions for the original problem are also met. \textit{A major challenge of the condensation is that its complexity may increases exponentially with the number of \acp{UAV}} \cite{4275017}. Therefore, we use a novel condensation whose complexity is linear, which can well cope with the massive \ac{UAV} case. Specifically, we first consider a simple case for the condensation of $1+a_{i}$, i.e., $S=1$, that satisfies requirement \textit{iii)}, it is easy to know that 
  \begin{equation}
    \begin{aligned}
    & w_{i} = \frac{a_{i,0}}{1+a_{i,0}} \quad 
    \end{aligned}.
    \label{eq:condensation}
    \end{equation} To further meet requirement \textit{ii)}, we have 

    \begin{equation}
      \begin{aligned}
      & c_{i} = (1+a_{i,0}) a_{i,0}^{-w_i} 
      \end{aligned}
      \label{eq:condensation}
      \end{equation} and it is straightforward to check that the requirement \textit{i)} is also met. Thus, for a general condensation of $g_i(\mathbf A)$ \addnew{at the approximation point $\mathbf A_0$}, $c_i$'s and $w_{is}$'s can be calculated as 
  \begin{equation}
    \begin{aligned}
    & w_{is} = \frac{a_{is,0}}{1+a_{is,0}}, \quad  \\ 
    & c_{i} = g_i(\mathbf A_0)(\prod_{s=1}^{S} a_{is,0}^{w_{is}} )^{-1} 
    \end{aligned}.
    \label{eq:condensation}
    \end{equation} It can be known that the complexity of the proposed condensation is linearly increasing with $S$, which is essential to make the \ac{SCA} feasible for the uplink communications of a large number of UAVs. 
    

    \begin{algorithm}[t]
      \setstretch{1}
      {\textbf {Input}: A feasible power allocation $\mathbf p$ for initialization\\} 
      {\textbf {Output}: Optimized $\mathbf p^{\ast}$ that maximizes the minimum \ac{SE}.}
      
      \begin{algorithmic}[1]
        \addnew{
      \STATE \textbf{Repeat:}
      \STATE Calculate $\mathbf A_0$ according to the current power allocation $\mathbf p$, i.e., $\mathbf A_0 = [\gamma_{is}(\mathbf p)]$.
      \STATE Perform condensation at the current point $\mathbf A_0$, i.e., calculate $c_i$'s and $w_{is}$'s at $\mathbf A_0$ according to \eqref{eq:condensation}.
      \STATE Solve the \ac{GP} problem with the condensation applied. The current power allocation $\mathbf p$ is updated as the newly optimized power allocation. 
      \STATE \textbf{Until} the power allocations obtained in two successive iterations change little, i.e., $||\mathbf p_{\text{now}} - \mathbf p_{\text{pre}} || < \epsilon$ with $\epsilon$ being a pre-defined tolerance. The final optimized power allocation $\mathbf p^{\ast}$ is then chosen as the current power allocation. 
       \caption{Solving problem \eqref{eq:fremaxmin_2}. \label{al1}}
       }
      \end{algorithmic}
      \end{algorithm}

\textbf{\textit{Considering the uplink SC constraint:}}
The power level of the same \ac{UE} at different bandwidth segments obtained by \aref{al1} can be arbitrarily different. This is applicable whenever the uplink \ac{SC} constraint is no longer needed. Considering the \ac{SC} constraint to maximize the minimum \ac{SE} is NP-hard \cite{5062197}. Thus we proposed a heuristic \aref{al2} for this purpose based on \aref{al1}. The proposed algorithm mainly includes three steps. \textit{i)} Obtain the first power allocation $\mathbf p^{\ast}$ using \aref{al1} without \ac{SC} constraint. \textit{ii)} Allocate continuous frequency resources for \ac{UAV}-\acp{UE} according to the result obtained in \textit{i)}. \textit{iii)} Perform \aref{al1} again considering the \ac{SC} constraint as posed by step \textit{ii)}. That is, for each \ac{UAV}, the power density on the allocated continuous segments is kept the same, and the power density of non-allocated segments is zero.

\begin{algorithm}[t]
  \setstretch{1}
  {\textbf {Input}: A feasible power allocation $\mathbf p$ for initialization\\} 
  {\textbf {Output}: Optimized $\mathbf p^{\ast}$ that maximizes the minimum \ac{SE} while also fulfilling \ac{SC} constraint.}
  
  \begin{algorithmic}[1]
  \STATE Obtain $\mathbf p^{\ast}$ using \aref{al1}.
  \STATE Allocate continuous frequency resources based on the \acp{SINR} achieved at $\mathbf p^{\ast}$ using \aref{al3}.
  \STATE Perform \aref{al1} again with the \ac{SC} constraint considered to update $\mathbf p^{\ast}$. 
   \caption{A heuristic algorithm solving problem \eqref{eq:fremaxmin_2} with \ac{SC} constraint. \label{al2}}
  \end{algorithmic}
  \end{algorithm}

  \begin{figure*}
    \begin{center}
      \stackunder[1pt]{\includegraphics[width=0.32\textwidth]{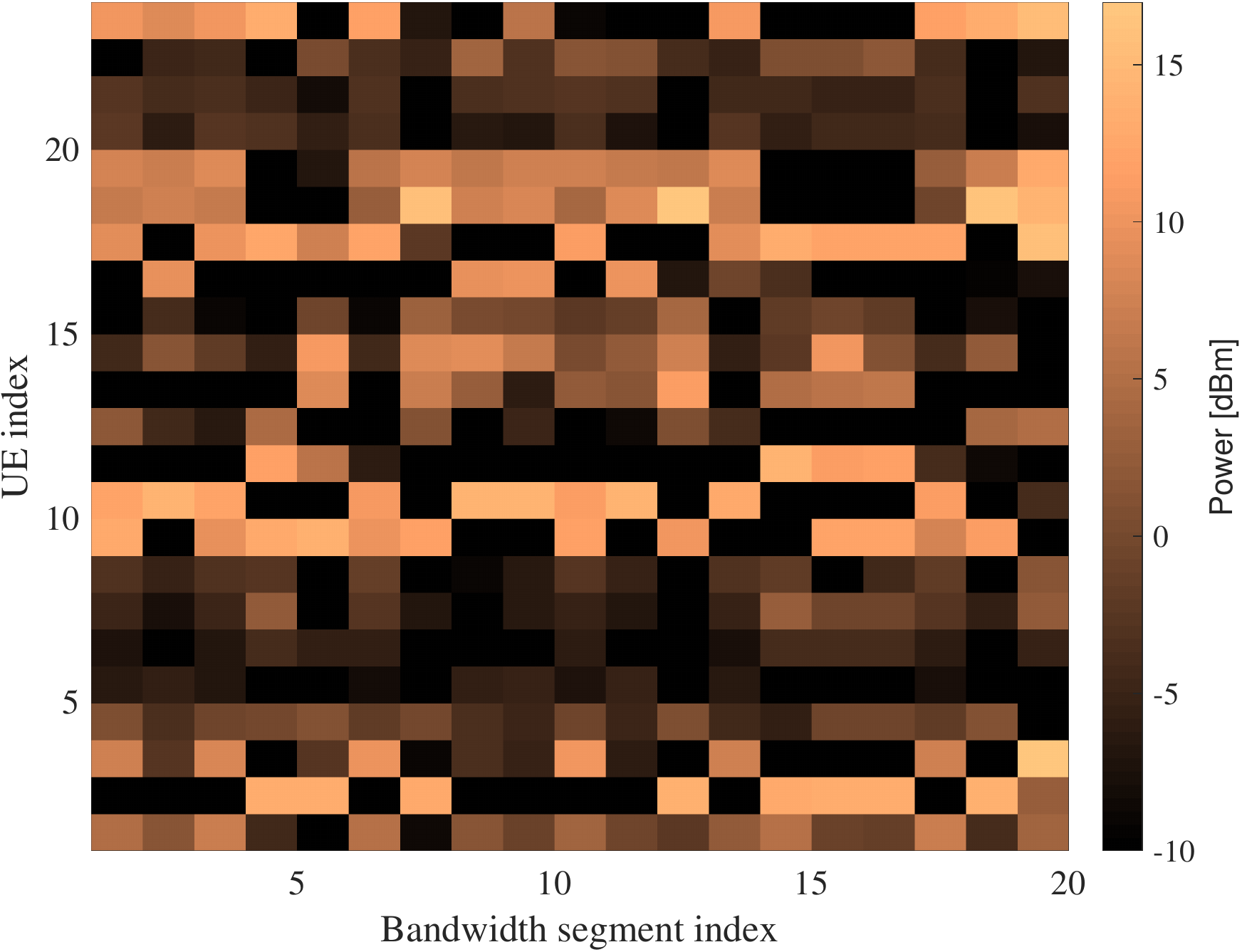}}{\small (a)}
      \stackunder[2.5pt]{\includegraphics[width=0.32\textwidth]{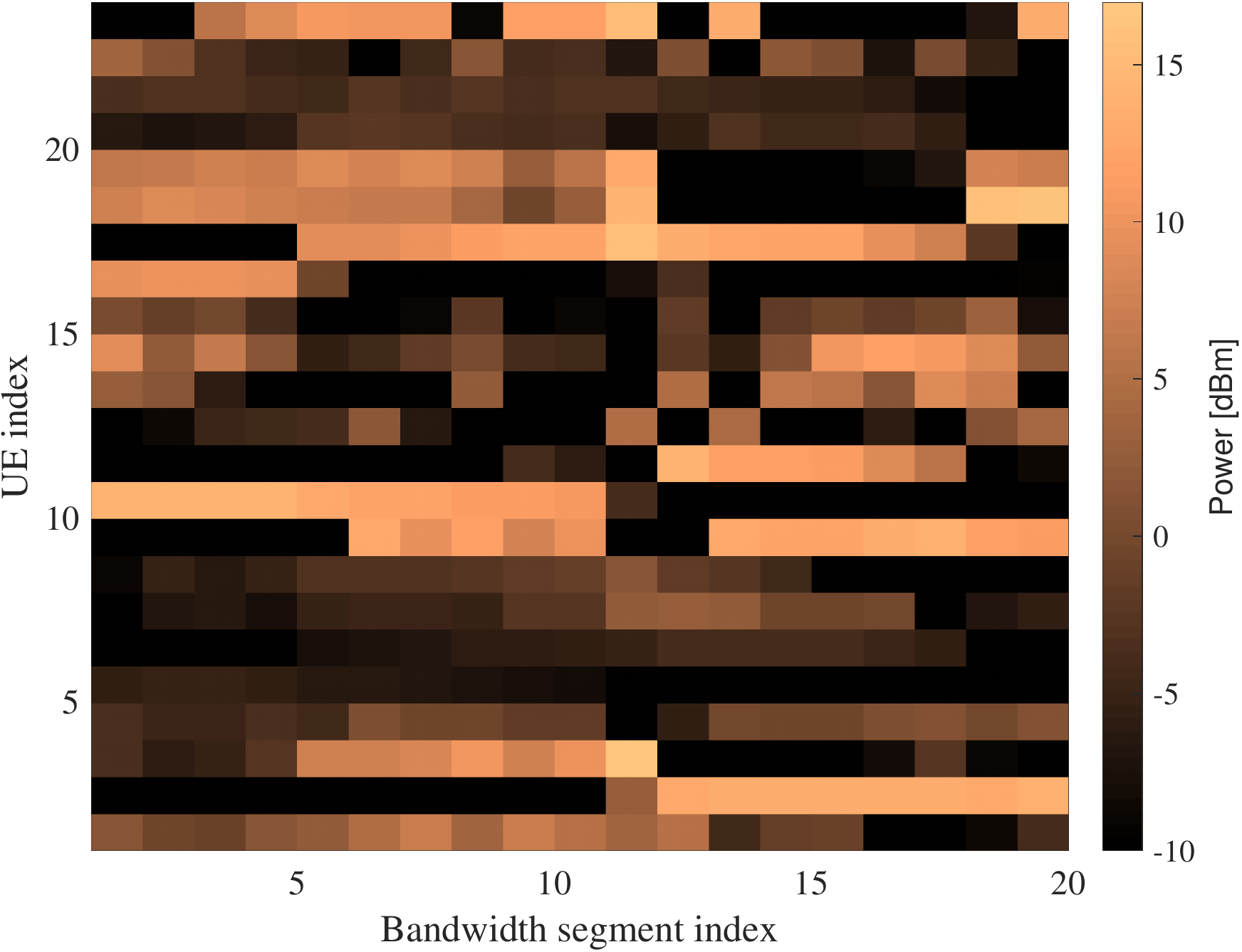}}{\small (c)}
      \stackunder[2.5pt]{\includegraphics[width=0.32\textwidth]{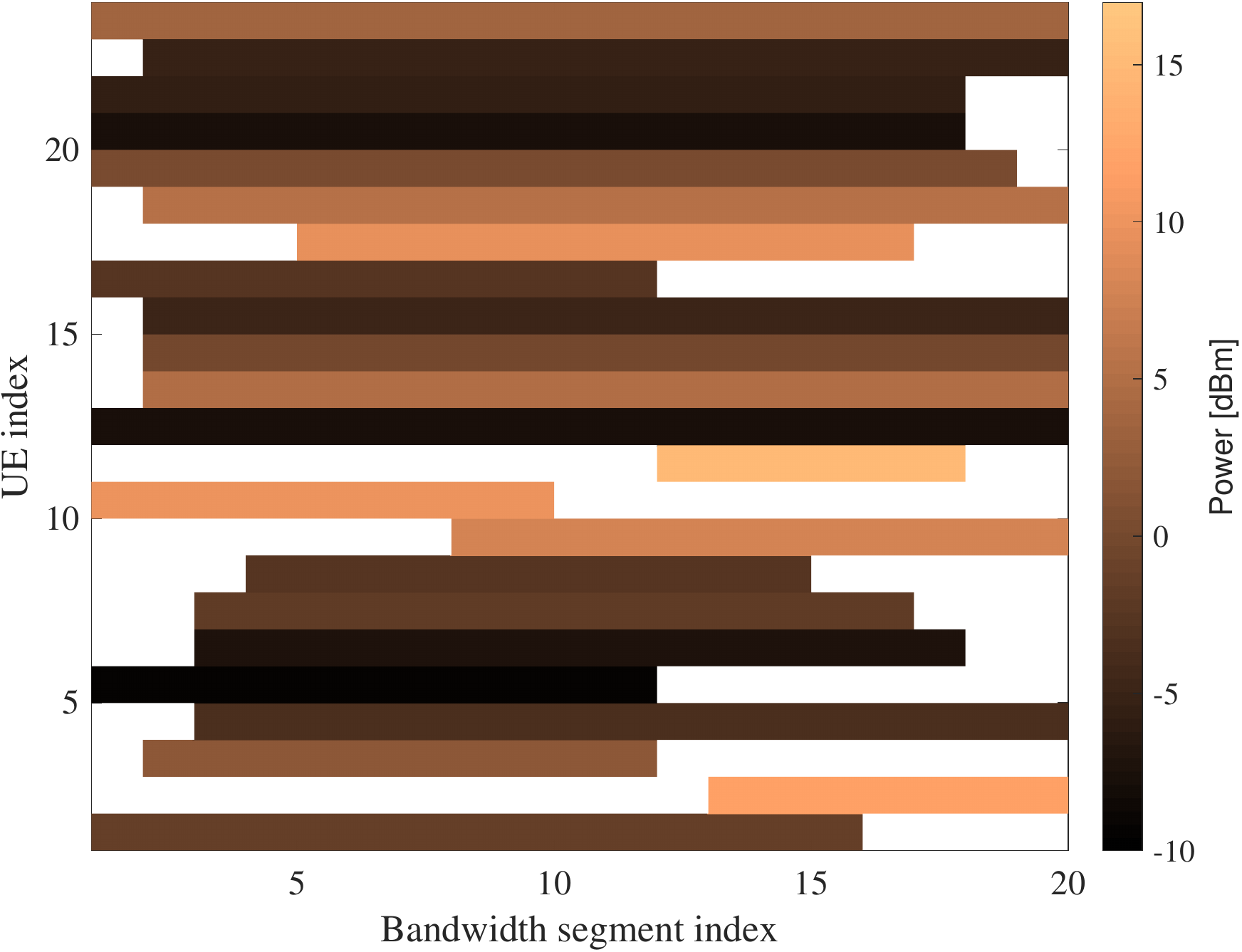}}{\small (e)}
      \stackunder[2.5pt]{\includegraphics[width=0.32\textwidth]{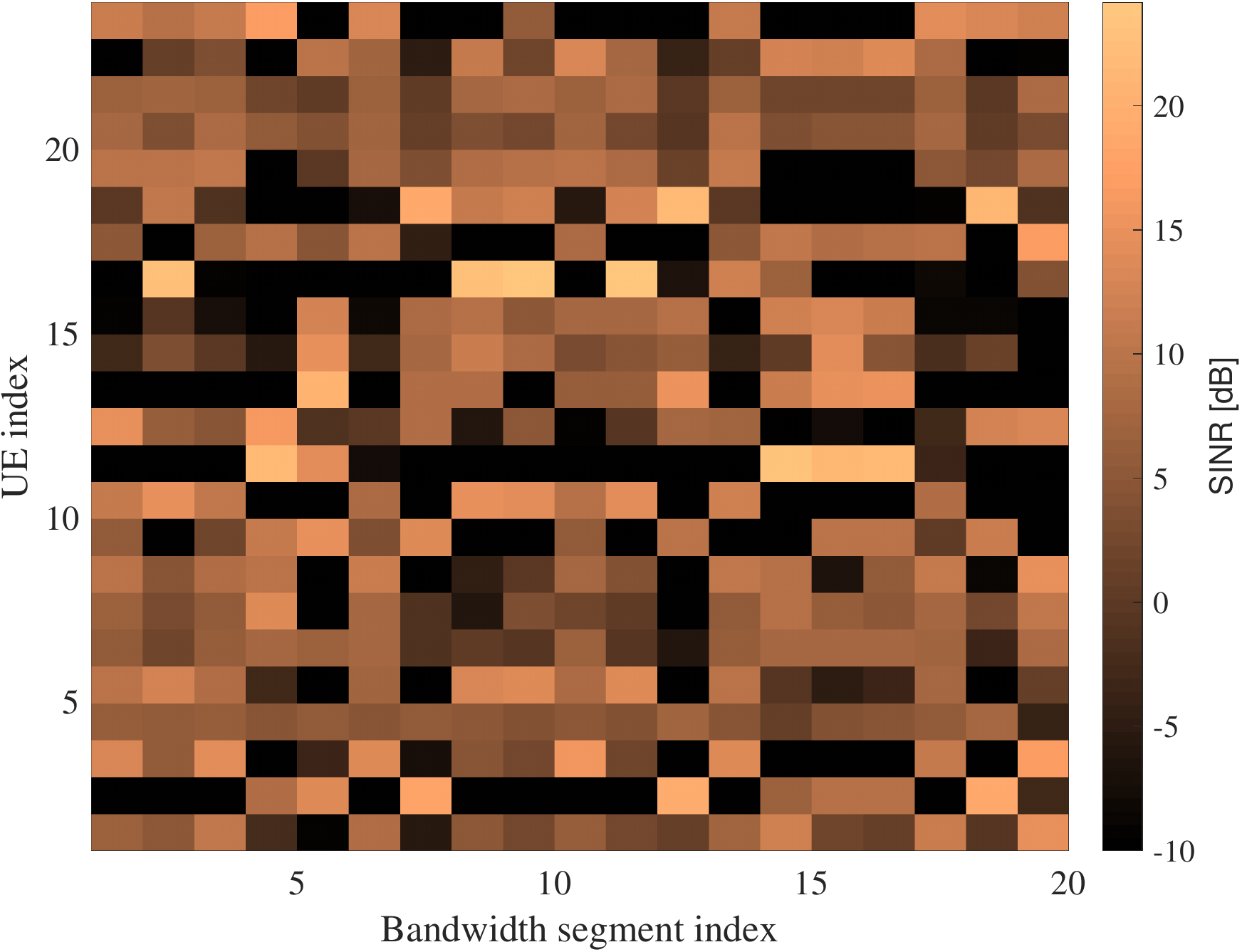}}{\small (b)}
      \stackunder[3pt]{\includegraphics[width=0.32\textwidth]{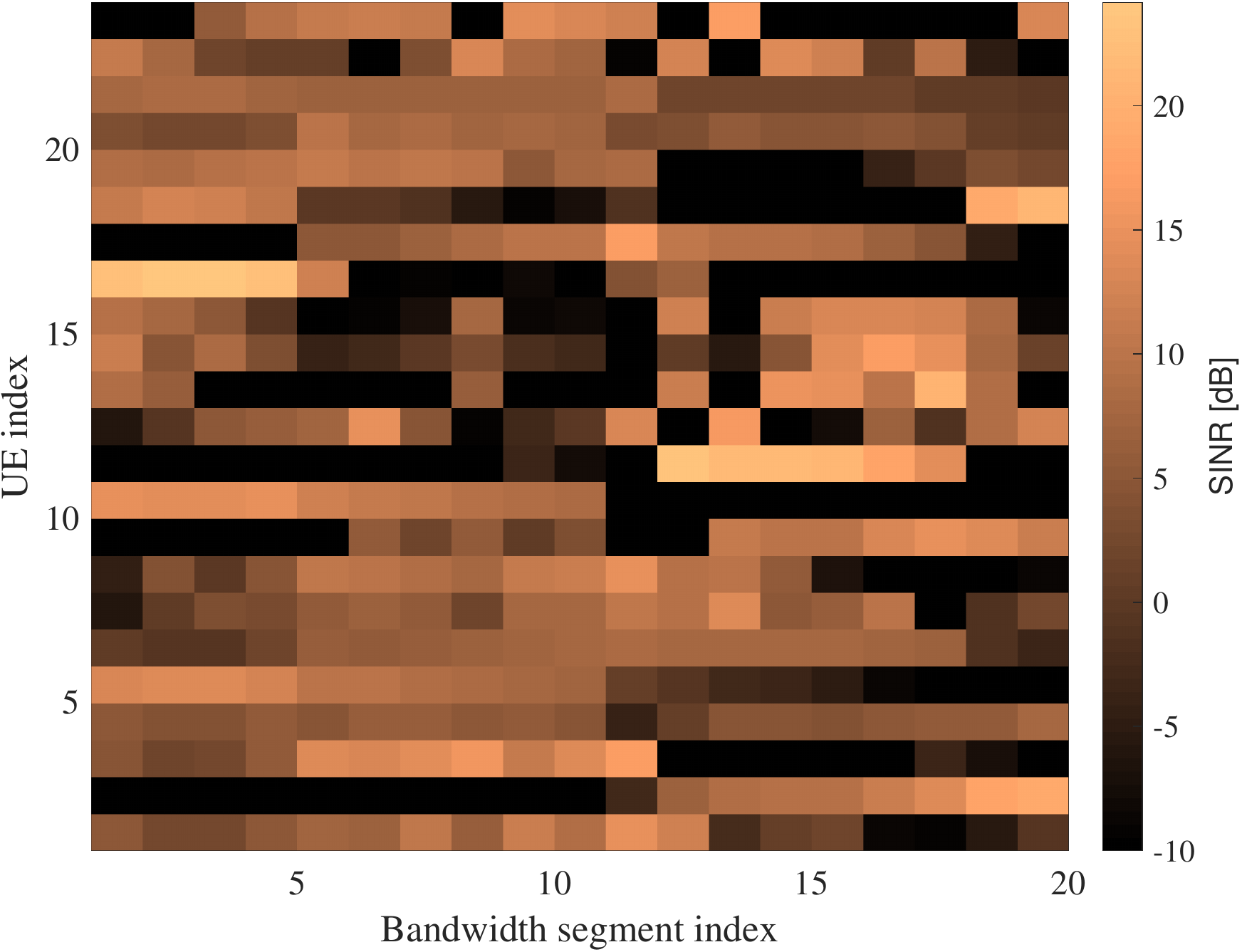}}{\small (d)}
      \stackunder[3pt]{\includegraphics[width=0.32\textwidth]{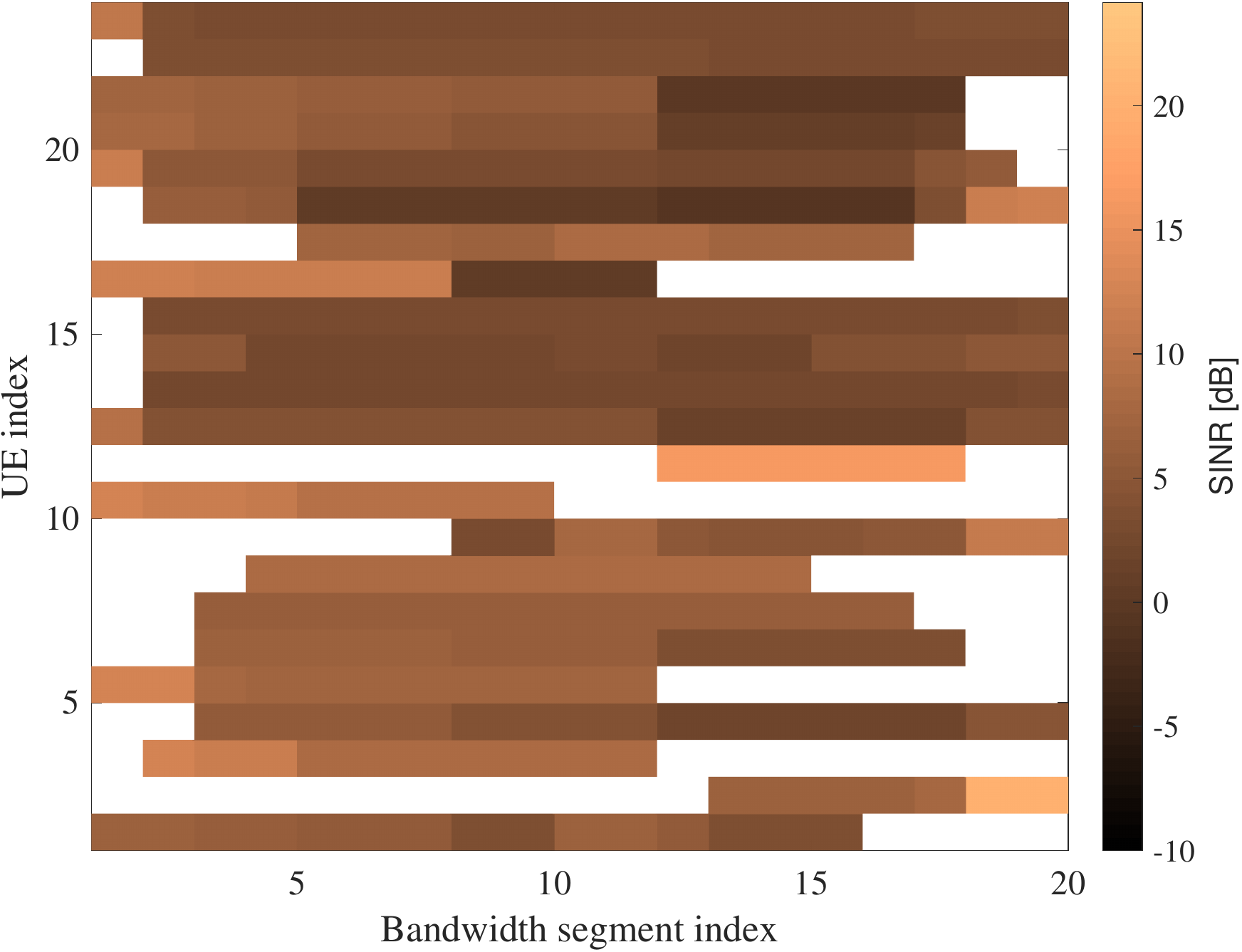}}{\small (f)}
    \end{center}
    \caption{An example application of \aref{al2}. (a) Power allocation obtained using \aref{al1}. (b) The corresponding \ac{SINR} matrix. (c) Reordered power allocation in \aref{al3}. (d) The corresponding reordered \ac{SINR} matrix. (e) Final power allocation with \ac{SC} constraint considered.  (f) The corresponding \ac{SINR} matrix of \ac{SC} transmission. \label{fig:scband}}
    \end{figure*}

  The heuristic part (step \textit{ii)}) of \aref{al2} is described as \aref{al3}. The purpose is to find continuous bandwidth segments of each \ac{UE} with powers that change as little as possible. \textit{Here, we exploit an important characteristic of the \ac{A2G} channel. That is, the channels are \addnew{close to frequency-flat, as shown in Fig.\,\ref{fig:channelmodels}(d)}. Therefore, we can reorder the columns of $\mathbf p^\ast$ obtained in step \textit{i)} while keeping the same optimization performance, which is essentially not possible for terrestrial \acp{UE}.} 
  Intuitively, if there exist two or more \ac{UAV}-\acp{UE} causing significant interference to each other on the same frequency resource, they probably will use different resources, and their \acp{SINR} on some segments could be very low. To increase the minimum \ac{SE} as possible, it is necessary to identify and put emphasis on this kind of \acp{UE}. We thus exploit the \ac{SINR} matrix to reorder columns. 
  The first step in \aref{al3} is to find the \ac{UE} with index $i$ that has the minimum $\gamma_{is}$ in the \ac{SINR} matrix $\mathbf \Gamma(\mathbf p^\ast)= [\gamma_{is}]$. Then the \ac{SINR} matrix and power matrix are updated by rearranging the columns according to descending/ascending order of the $S$ \acp{SINR} of the $i$th \ac{UAV}-\ac{UE} on the $S$ bandwidth segments. This is trying to smooth the \ac{SINR} change (and power change) at neighboring bandwidth segments.  Finally, for each UE, expanding continuous frequency resources starting with the bandwidth segment having the largest \ac{SINR} till the summed rate on these bandwidth segments is larger than a pre-defined threshold of the original overall summed rate on all $S$ segments. \fref{fig:scband} illustrates an example application of \aref{al2}. In this example, 24 cells are active each with a \ac{UAV}-\ac{UE}. Bandwidth $B$ is divided into $S= 20$ segments. It can be clearly observed from \fref{fig:scband}(c) or (d) that \acp{UE}\,\#2 and \#10 are using interleaved bandwidth segments, which means that they cause severe interference to each other if transmitting on the same frequency resource, e.g. when $S=1$. The finally maximized minimum equivalent \ac{SINR} (i.e., $2^{SE}-1$) obtained using \aref{al2} in this case is 3.2\,dB. As a comparison, the maximized minimum equivalent \ac{SINR} obtained using \aref{al1} without considering \ac{SC} constraint, as shown in \fref{fig:scband}(b), is 4.9\,dB due to more freedom in power allocation. Whereas when using the whole bandwidth, i.e. $S=1$, for \ac{SC} transmission, the maximized minimum equivalent \ac{SINR} is only -3.6\,dB. This significant gain of around 7\,dB in \ac{SINR} illustrates the effectiveness of the proposed algorithm.

  \begin{algorithm}[t]
    \setstretch{1}
    {\textbf {Input}: SINR matrix $\mathbf \Gamma =[\gamma_{is}]$ achieved at final $\mathbf p^\ast$ using \aref{al1} without considering \ac{SC} constraint. \\} 
    {\textbf {Output}: Bandwidth allocation fulfiling SC constraint.}

    \begin{algorithmic}[1]
    \STATE Find the index $i$ of the \ac{UE} having the minimum \ac{SINR}, i.e., $(i,s) = \arg \min_{j,v} \gamma_{jv}$.
    \STATE Update $\mathbf \Gamma$ by rearranging its columns according to descending order of the \acp{SINR} at $S$ bandwidth segments of the $i$th \ac{UE}.
    \STATE For each UE, start with its bandwidth segment with largest \ac{SINR} and expand continuously to neighboring ones till the summed rate exceeds a pre-defined proportion $\eta$ of the original overall rate.
     \caption{A heuristic algorithm for \ac{SC} bandwidth allocation. \label{al3}}
    \end{algorithmic}
    \end{algorithm}

\subsection{Time-domain maximization of the minimum \ac{SE} of \ac{UAV}-\acp{UE}\label{sec:tdmaxmin}}

This is an algorithm proposed for the case where the \ac{UAV}-\acp{UE} are allowed to transmit for a certain period, e.g., a set $\mathcal{T} = \{1,\cdots,T\}$ of \acp{TTI}, which is still within the coherent time of the channels. In each \ac{TTI}, we assume a \ac{UAV}-{UE} is using the whole $B$ with the same power density to not complicate the expression without losing the essence. The problem can then be formulated as 
\begin{equation}
  \begin{aligned}
    \addnew{\underset{\mathbf p}{\text{maximize}}} \quad  & \min_i R_i \\
  \text{subject to} \quad & p_{i,\text{min}} \leq p_{it} \leq  p_{i,\text{max}}, \forall i \in \mathcal{N},  \forall t \in \mathcal{T}
  \end{aligned}
  \label{eq:timemaxmin}
  \end{equation} where $\mathbf p = [p_{it}] $ with $p_{it}$ indicating the output power of the $i$th \ac{UAV}-\ac{UE} at the $t$th \ac{TTI}. Moreover, $R_i$ is also modified as 
  \begin{equation}
    \begin{aligned}
    R_{i}(\mathbf{p}) = T^{-1} \sum_{t\in \mathcal{T}} \log_2(1+\gamma_{it} )
    \end{aligned}
    \end{equation} 
  with
  \begin{equation}
    \begin{aligned}
      \gamma_{it}(\mathbf{p}) = \frac{p_{it} G_{ii}}{N_0 B + \sum_{j\in \mathcal{N}, j\neq i} p_{jt} G_{ji} } 
    \end{aligned}.
    \end{equation} This problem, with a similar format to \eqref{eq:fremaxmin_2}, can be solved essentially almost the same as described in \aref{al1}. \textit{Nevertheless, it is not exactly the same to \eqref{eq:fremaxmin_2}.} One reason is that power levels of a \ac{UAV}-\ac{UE} at different \acp{TTI} can be arbitrarily different, i.e., the \ac{SC} constraint is naturally met. The other reason is that in \eqref{eq:fremaxmin_2}, since a \ac{UAV}-\ac{UE} can use less bandwidth, its maximum power density is higher than that achieved in \eqref{eq:timemaxmin}, which may result in different performances. They are also the reasons that we discuss time-domain and frequency-domain algorithms separately. 
    

\subsection{Frequency-domain maximization of the sum \ac{SE} of \ac{UAV}-\acp{UE}\label{sec:fdmaxsum}}  
Besides maximizing the minimum \ac{SE} of \acp{UAV} in the system, another principle can be to maximize the overall \ac{SE} of all \acp{UAV} in the system, depending on the application scenarios. Based on the notations as described in \sref{sec:fdmaxmin}, we formulate the problem as
\begin{equation}   
\begin{aligned}
  \addnew{\underset{\mathbf p}{\text{maximize}}} \quad  & \sum_{i=1}^{N} R_i \\ 
  \text{subject to} \quad & p_{i,\text{min}} \leq p_i \leq  p_{i,\text{max}}, \forall i \in \mathcal{N} \\
    & p_i = \sum_{s\in\mathcal{S}} p_{is}, \forall i \in \mathcal{N} \\
    & R_i \geq  R_{i,\text{min}}, \forall i \in \mathcal{N} 
  \end{aligned}
  \label{eq:fremaxsum}
  \end{equation} where the last line in \eqref{eq:fremaxsum} is the \ac{QoS} constraints. We propose \aref{al4} to solve \eqref{eq:fremaxsum} with the \ac{SC} constraint also considered, which can be the most complicated case. The key is to find an initial feasible \ac{SC} power allocation (may not exist) that meets the \ac{QoS} constraint, which must be based on the algorithm proposed in \sref{sec:fdmaxmin}. We do not redundantly write algorithms for other easier cases. For example, without \ac{SC} constraint, the step\,{\small 1} in \aref{al4} can be modified as to exploit \aref{al1} to find an initial feasible power allocation, and the \ac{SC} constraint is also removed from step\,{\small 3}.

  \begin{algorithm}
    \setstretch{1}
    {\textbf {Output}: Optimized $\mathbf p^{\ast}$ that maximizes the sum \ac{SE} fulfilling both \ac{QoS} and \ac{SC} constraints.}
    \begin{algorithmic}[1]
    \STATE Apply \aref{al2} to obtain the uplink \ac{SC} bandwidth allocation and the maximized-minimum \ac{QoS}.
    \IF{the achieved \ac{QoS} can meet the constraint in \eqref{eq:fremaxsum} }
    \STATE  Use the power allocation obtained in step {\small{1}} as a feasible initialization, and perform a series of \acp{GP} considering \ac{SC} bandwidth allocation till convergence. 
  \ELSE
    \STATE No feasible solution.
  \ENDIF
     \caption{Solving problem \eqref{eq:fremaxsum} considering \ac{SC} constraint. \label{al4}}
    \end{algorithmic}
    \end{algorithm}

 \subsection{Time-domain maximization of the sum \ac{SE} of \ac{UAV}-\acp{UE}} \label{sect:tdmaxsum}

 This can be formulated from \sref{sec:tdmaxmin}, as similarly done from \sref{sec:fdmaxmin} to \sref{sec:fdmaxsum}. We thus omit the detailed formulas.
 

\subsection{Scheduling and power control for multiple \ac{UAV}-\acp{UE}}\label{sect:twoue}

Multiple \ac{UAV}-\acp{UE} in a cell, e.g. a set $\mathcal{K}_i = \{1,\cdots,K_i\}$ of \acp{UE} with the $K_i$ highest time-domain \ac{PF} priorities in the $i$th cell, may be allowed to transmit simultaneously. The methodology is essentially similar to what we have discussed for single-\ac{UE} cases. For example, to maximize the minimum \ac{SE} of all \ac{UAV}-\acp{UE} in the network, the problem can be formatted similarly to \eqref{eq:fremaxmin} as
\begin{equation}
  \begin{aligned}
    \addnew{\underset{\mathbf p}{\text{maximize}}}\quad  & \min_{i,k_i} R_{i,k_i} \\ 
  \text{subject to} \quad & p_{i,k_i,\text{min}} \leq p_{i,k_i} \leq  p_{i,k_i,\text{max}},  & \forall i,  k_i\\
    & p_{i,k_i} = \sum_{s\in\mathcal{S}} p_{i,k_i,s},  & \forall i, k_i 
  \end{aligned}
  \label{eq:multiUEmaxmin}
  \end{equation}
  with the \ac{SE} of the $k_i$th \ac{UE} in the $i$th cell modified as 
  \begin{equation}
    \begin{aligned}
    R_{i,k_i}(\mathbf{p}) = K_iS^{-1} \sum_{s\in \mathcal{S}} R_{i,k_i,s} 
    \end{aligned}.
    \end{equation}  
  because in total $K_i$ \ac{UAV}-\acp{UE} share the frequency resources in the $i$th cell. With the \ac{SC} constraint further considered, one can exploit \aref{al2} to conduct packet scheduling and power allocation for individual \acp{UE}. An additional consideration is that \acp{UE} in the same cell have to use orthogonal bandwidth segments unless \ac{NOMA} technique is applied. 
  
  
  





\section{Simulation and discussions\label{sect:simulations}}\label{sect4}

In this section, we demonstrate the performances of the different algorithms proposed in \sref{sec:gp} via extensive simulations. Full-buffer mode and bursty-traffic mode are considered in \sref{sect:full} and \sref{sect:bursty}, respectively. To ease the description of the proposed algorithms, the abbreviations as shown in \tref{tab:algorithmname} are applied. We use e.g. ``FD\,SC\,Max-Min'' to denote \textit{frequency domain maximization of the minimum \ac{SE} considering \ac{SC} constraint}. The abbreviation style is similarly applied to other algorithms.

\subsection{full-buffer mode}\label{sect:full}

\textit{1) The case of single UE per cell:} 

In this subsection, we focus on the full-buffer uplink transmission mode, where \ac{UAV}-\acp{UE} are assumed to constantly transmit uplink data. In the simulation, a network with 48 sectored cells as illustrated in \fref{fig:cells} is applied. Important simulation parameters are the same as included in \tref{tab:sim_parameter} in \sref{sect:intferences}.    
We assume that an array consisting of six antennas with 60$^\circ$ \acp{HPBW} is onboard the \ac{UAV} \cite{9082692}, and the best directional antenna is switched on for the uplink communication. 
The percentage of active cells ranges from 0.5 to 1. That is, in each realization, a random portion of the 48 cells are active with a \ac{UAV}-\ac{UE} in full-buffer mode transmission. Totally 300 realizations, each with random locations of \acp{UAV}, are performed. The first consideration is to obtain the statistical performance of the proposed algorithms. The second is to include the effect of mobility of \acp{UAV} which can statistically lead to randomized locations of \acp{UAV} in each cell. 

\begin{figure}
  
  \begin{center}
    
    \redfbox{\subfigure[]{\includegraphics[width=0.45\textwidth]{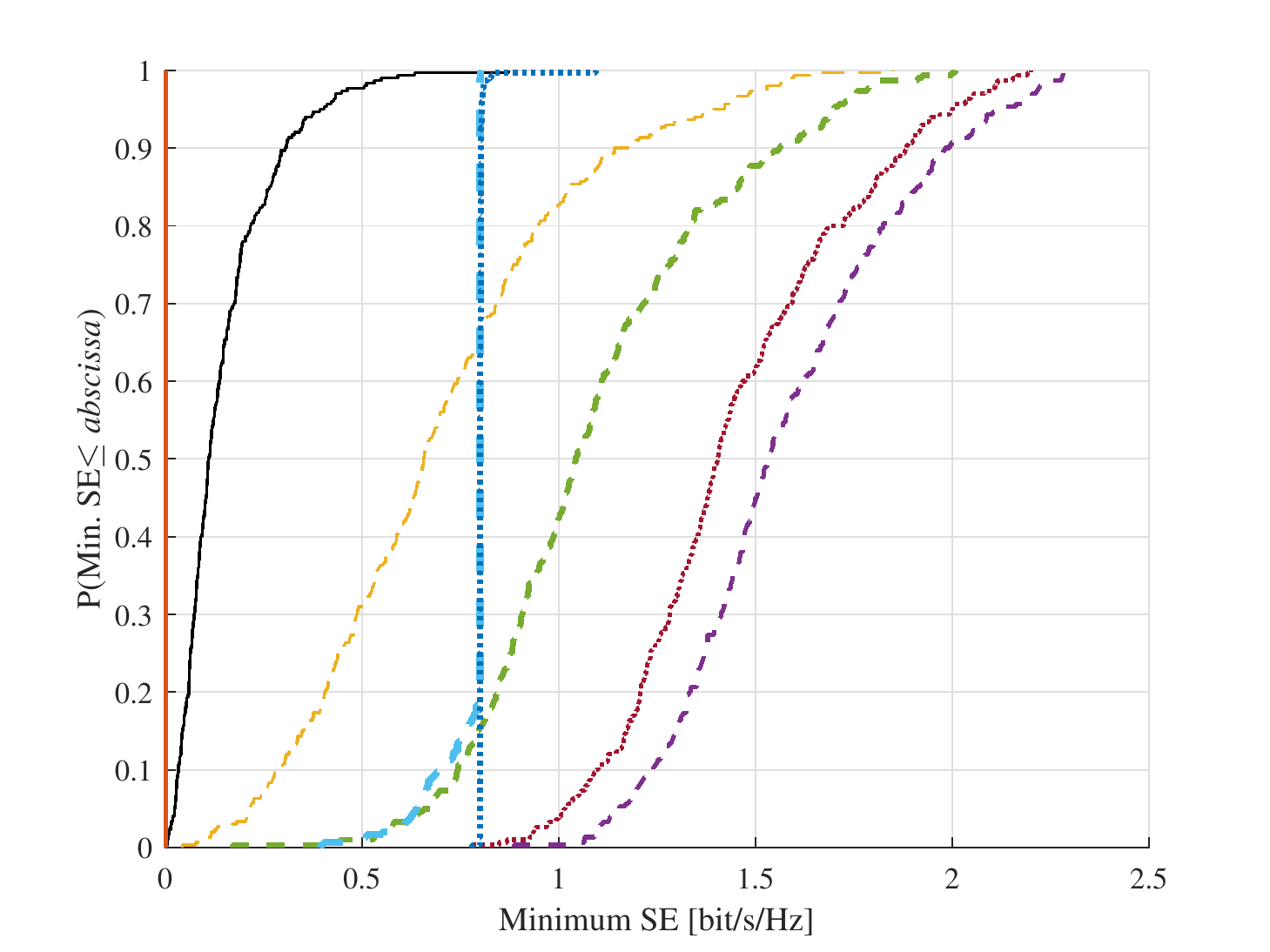}{\psfrag{F(x)}[l][l][0.65]{Omnidirectional}}}}
    \redfbox{\subfigure[]{\includegraphics[width=0.45\textwidth]{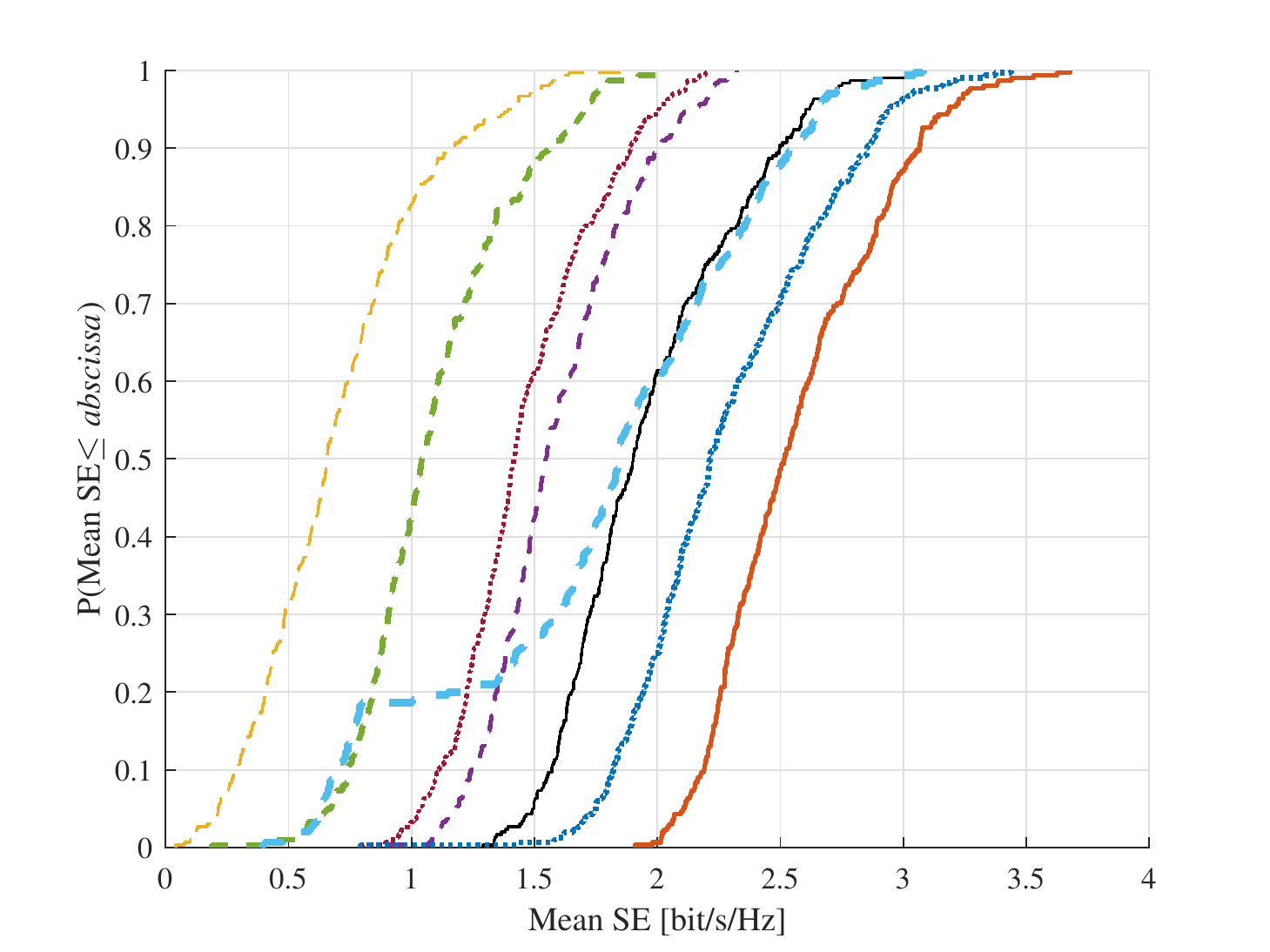}}}
    \redfbox{\subfigure[]{\includegraphics[width=0.45\textwidth]{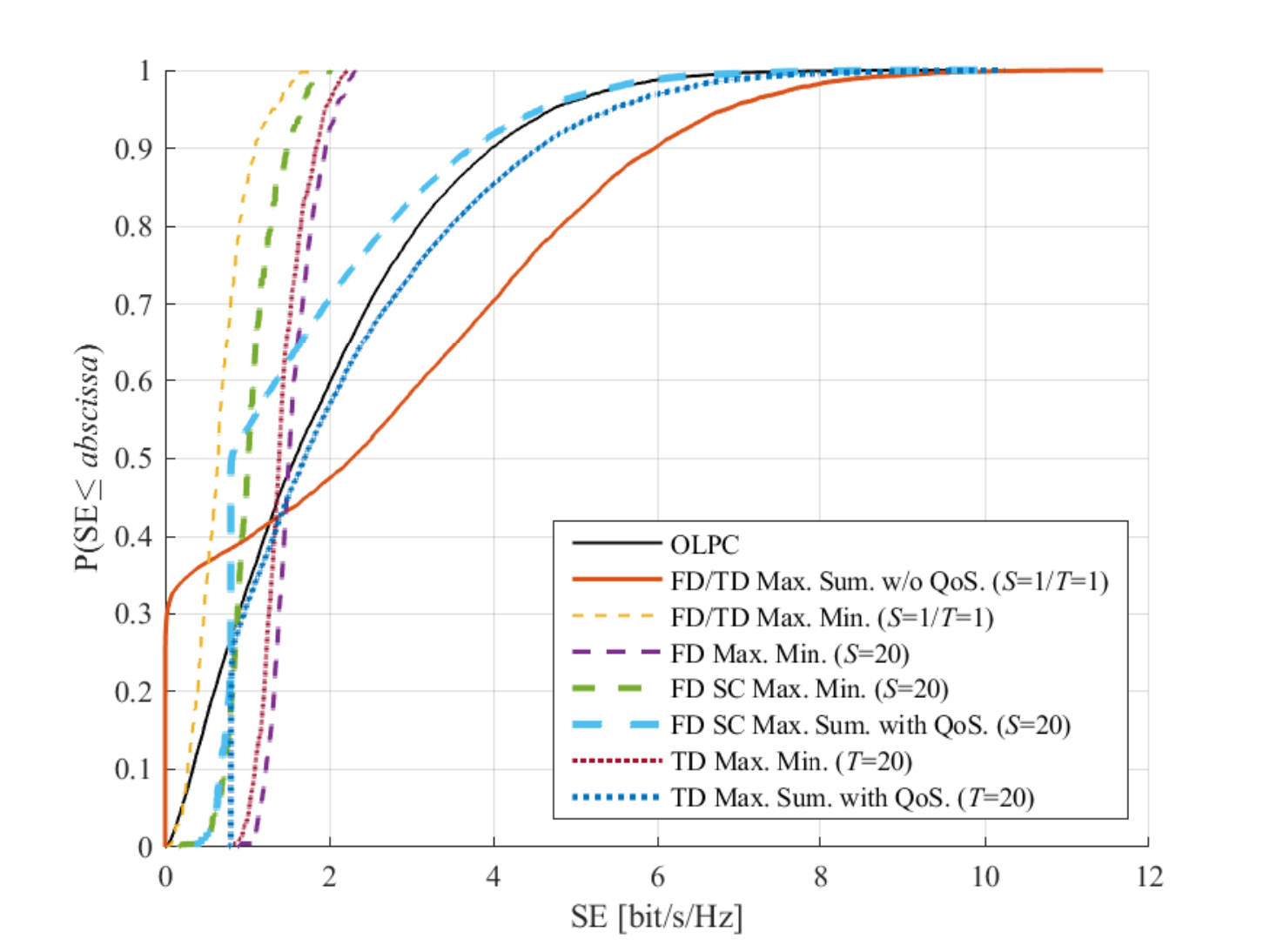}}}
  \end{center}
  
  \caption{\acp{CDF} of \acp{SE} of \acp{UAV} in full-buffer mode. (a) Minimum \acp{SE} obtained in individual realizations. (b) Mean \acp{SE} obtained in individual realizations. (c) \acp{SE} of all \ac{UAV}-\acp{UE} in all realizations.\label{fig:fullbuffer}}
  \end{figure}



  \begin{table}[]
    \centering
    \caption{Algorithms utilized in the simulation.}
    \label{tab:my-table}
    \scalebox{0.8}{
    \begin{tabular}{lc}
    \hline\hline
   {Abbreviation} & Referring to \\ \hline
                           OLPC                                     & \eqref{eq:olpc} in \sref{sect:intferences}   \\ 
                           FD/TD Max-Sum w/o QoS ($S$=1/$T$=1)    &  \sref{sec:fdmaxsum}  \\  
                           FD/TD Max-Min ($S$=1/$T$=1)              &   \aref{al1} in \sref{sec:fdmaxmin}      \\ 
                           FD Max-Min ($S$=20)                      &  \aref{al1} in \sref{sec:fdmaxmin}     \\ 
                           FD\,SC\,Max-Min ($S$=20)                 & \aref{al2} in \sref{sec:fdmaxmin} \\ 
                           FD\,SC\,Max-Sum with QoS ($S$=20)         & \aref{al4} in \sref{sec:fdmaxsum}    \\ 
                           TD Max-Min ($T$=20)                  &  \sref{sec:tdmaxmin}    \\
                           TD Max-Sum with QoS ($T$=20)          &  \sref{sect:tdmaxsum}    \\ \hline\hline 
    \end{tabular}}
    \label{tab:algorithmname}

    \vspace{0.3cm}

    {\small 
    FD: Frequency domain; 
    TD: Time domain; 
    Max-Sum: Maximization of the sum SE; 
    Max-Min: Maximization of the minimum SE;
    QoS: QoS constraint; 
    SC: Single carrier; 
     $S$: Segment number in frequency domain;
    $T$: TTI number. }

    \end{table}

\begin{table}[]
    \centering
    \caption{Performance gains (in percentage) for the $H$-th percentile \acp{SE} of all \ac{UAV}-\acp{UE} in the full buffer mode compared to \ac{OLPC}. The top three best algorithms are bolded.  
    }
    \label{tab:my-table}
    \redfbox{
    \scalebox{0.75}{
    \begin{tabular}{l|r|r|r|r}
    \hline
   {\diagbox{Algorithms}{$H$}} & {10} & {20} & {50} & {100} \\ \hline
                           FD/TD Max-Sum w/o QoS. ($S$=1/$T$=1)    &   -100\%  &   -100\%  &   \textbf{44\%}  &   \textbf{16\%}  \\ \hline 
                           FD/TD Max-Min ($S$=1/$T$=1) &   -18\%  &   -33\%  &   -59\%  &   -81\%  \\ \hline 
                           FD Max-Min ($S$=20)&   \textbf{265\%}  &  \textbf{123\%}  &   \textbf{-5\%}  &   -76\%  \\ \hline 
                           FD SC Max-Min ($S$=20)&   119\%  &   \textbf{38\%}  &   -36\%  &   -80\%  \\ \hline 
                           FD SC Max-Sum with QoS. ($S$=20)&   99\%  &   32\%  &   -49\%  &   \textbf{1\%}  \\ \hline 
                           TD Max-Min ($T$=20)&  \textbf{225\%}  &   \textbf{102\%}  &   -12\%  &   -77\%  \\ \hline 
                           TD Max-Sum with QoS. ($T$=20)&   \textbf{138\%}  &   35\%  &   \textbf{7\%}  &   \textbf{4\%}  \\ \hline 
    \end{tabular}}
    }
    \label{tab:full_buffer_mode}
    \end{table}

  \fref{fig:fullbuffer} illustrates the \acp{CDF} of \acp{SE} obtained using the proposed algorithms as summarized in \tref{tab:algorithmname}.\footnote{\addnew{Although the closed loop power control (CLPC) \cite{5198853,3GPP36213} could have better performance for cell-edge UEs, it has many more parameters to be tuned compared to the OLPC. Moreover, it is still a non-optimized power control scheme. Therefore, we choose the tuned OLPC in the simulation as the non-optimized baseline for comparison. }} 
Specifically, \fref{fig:fullbuffer}(a) illustrates the \acp{CDF} of the minimum \acp{SE} achieved in individual realizations, \fref{fig:fullbuffer}(b) illustrates the \acp{CDF} of the mean \acp{SE} obtained in individual realizations, and in \fref{fig:fullbuffer}(c) the \acp{CDF} of \acp{SE} of all \ac{UAV}-\acp{UE} in all realizations are presented. In the \ac{OLPC} algorithm, \ac{UAV}-\acp{UE} are transmitting on the whole reserved bandwidth according to \eqref{eq:olpc} with the optimized $P_0$ and $\alpha$ as shown in \tref{tab:sim_parameter}. For other algorithms, Max-Sum is conducted without QoS constraint. In the frequency domain, three different Max-Min algorithms are performed which include FD Max-Min ($S=1$), FD Max-Min ($S=20$) and FD\,SC\,Max-Min ($S=20$). Moreover, FD\,SC\,Max-Sum ($S=20$) is also applied with \ac{QoS} constraint set as 0.8\,bit/s/Hz for all \ac{UAV}-\acp{UE}. Similarly in the time domain, TD Max-Min ($T=20$) and TD Max-Sum ($T=20$) with \ac{QoS} constraint as 0.8\,bit/s/Hz are conducted. \tref{tab:full_buffer_mode} summarizes the performance gains of these algorithms applied in the full-buffer mode for different groups of \ac{UAV}-\acp{UE} compared to that of \ac{OLPC}. According to the three subfigures of \fref{fig:fullbuffer} and \tref{tab:full_buffer_mode}, we have the following observations and findings.
\begin{itemize}

  \item The Max-Min algorithms can achieve better minimum \acp{SE} compared to that of \ac{OLPC} and Max-Sum without \ac{QoS} constraint. Especially, the minimum \ac{SE} of Max-Sum without \ac{QoS} constraint is almost always zero, which means that in the interference dominant scenario, there are always \ac{UE}(s) being sacrificed to maximize the sum \ac{SE}. Moreover, it can be observed that with a larger $S$ or no \ac{SC} constraint, FD Max-Min can achieve a larger minimum \ac{SE}. This is reasonable since a larger $S$ and/or no \ac{SC} constraint can provide more freedom for \ac{UAV}-\acp{UE} to avoid severe interference by transmitting on interleaved bandwidth segments. Furthermore, we can observe that the performance of TD Max-Min ($T=20$) is close to that of FD Max-Min ($S=20$) regarding the achieved minimum \ac{SE}. This is understandable because the \ac{SC} constraint is not a concern in the time domain while still keeping the similar freedom of avoiding severe interference by transmitting at interleaved times. Thus, TD Max-Min is considered a good option. Nevertheless, the power density in FD Max-Min can be higher which is advantageous for edge \acp{UE}, so that the performance of TD Max-Min is slightly lower than that of FD Max-Min

    \item Although the various TD/FD Max-Min algorithms can achieve better minimum \acp{SE}, they have less mean \acp{SE} compared to that of \ac{OLPC}. The reason is that in the Max-Min algorithms, all the \ac{UAV}-\acp{UE} in a realization are with the same \ac{SE}, as we can observe that the curves of Max-Min algorithms in \frefp{fig:fullbuffer}(a)-(c) are the same. This is also easy to be verified by contradiction as follows. Assume that there are one or several \acp{UE} with higher \acp{SE}, then their transmitting powers can be decreased to increase the minimum \ac{SE} until all \ac{UAV}-\acp{UE} have the same \ac{SE}. Therefore, the \acp{SE} of \acp{UE} in good conditions are limited leading to smaller mean \ac{SE}. In addition, it is straightforward that the Max-Sum without \ac{QoS} constraint can achieve the best mean \ac{SE}.
    
    \item Compared to \ac{OLPC}, Max-Min algorithms are favorable to \acp{UE} in bad conditions however limit the \acp{SE} of \acp{UE} in good conditions; whereas the Max-Sum without \ac{QoS} constraint sacrifices the \acp{UE} in bad conditions. By introducing \ac{QoS} constraints into Max-Sum, the compromise can be tuned. For example, it can be observed from \fref{fig:fullbuffer}(a) that the minimum \acp{SE} of FD\,SC\,Max-Sum with \ac{QoS} constraint and TD Max-Sum with \ac{QoS} constraint are realized as 0.8\,bit/s/Hz if there exist feasible solutions, otherwise the original Max-Min \acp{SE} are kept. It can be observed from \fref{fig:fullbuffer}(b) that the mean \acp{SE} are increased compared to that of the corresponding Max-Min algorithms. Moreover in \fref{fig:fullbuffer}(c), it can be observed that the \acp{SE} of \acp{UE} in bad conditions (e.g., \acp{UE} with \acp{SE} below the 20th percentile) are guaranteed, and the \acp{SE} of other \acp{UE} in better conditions are not limited too much either. It is worth noting that the TD Max-Sum with \ac{QoS} constraint herein can achieve overall better performance for both types of \acp{UE} compared to the \ac{OLPC} algorithm. 
  \end{itemize}

\textit{2) The case of multiple UEs per cell:}

\begin{figure}
  \begin{center}
  \redfbox{\includegraphics[width=0.45\textwidth]{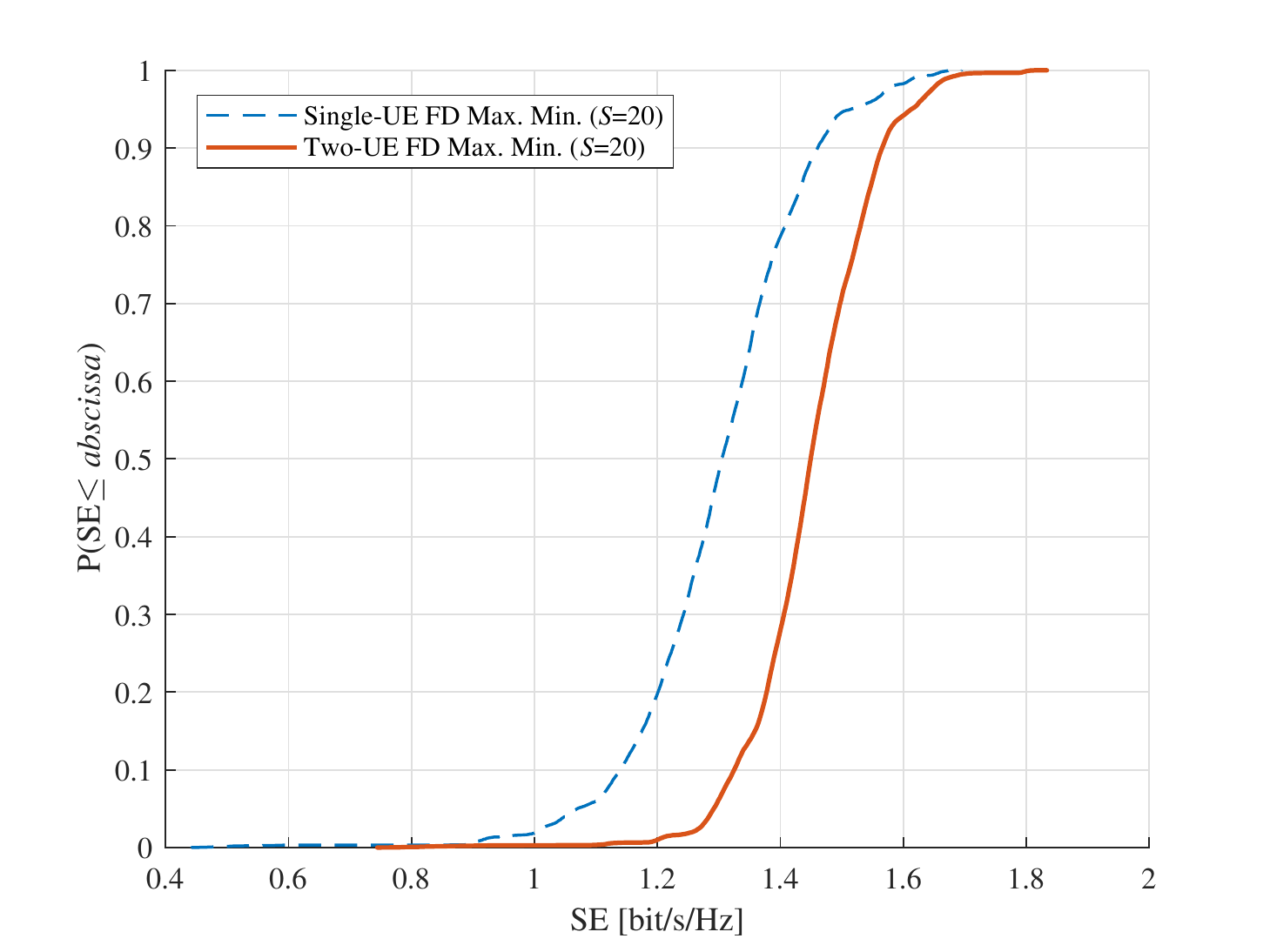}}
  \end{center}
  \caption{CDFs of \acp{SE} of \ac{UAV}-\acp{UE} obtained in the two-\ac{UE} scheduling and single-\ac{UE} scheduling in the full-buffer mode. \label{fig:twoue}}
  \end{figure} 

As discussed in \sref{sect:twoue}, it is possible that multiple \ac{UAV}-\acp{UE} can be scheduled in each cell. We consider herein a simple case, i.e. scheduling two \acp{UE} per cell in the full-buffer mode. All 48 cells are active, and the FD Max-Min algorithm with $S=20$ is utilized. \fref{fig:twoue} illustrates the \acp{SE} of two-\ac{UE} scheduling and the \acp{SE} of single-\ac{UE} scheduling. It can be observed that the performance of two-\ac{UE} scheduling is better than that of the single-\ac{UE} scheduling. This is due to that by scheduling more \acp{UAV}, each \ac{UAV} tends to use less bandwidth compared to that of single-UE scheduling. With the same allowed maximum transmit power, a higher maximum power density tends to be achieved in multiple-UE scheduling.  
For those \acp{UAV} with large path loss, i.e. power limited \acp{UAV}, higher power density is favorable for them to increase their \acp{SE}. This shows the potential of multiple-\ac{UE} scheduling. In the bursty-traffic mode in the sequel, we still focus on the single \ac{UE} scheduling. 




\subsection{bursty-traffic mode}\label{sect:bursty}

In this subsection, we focus on the bursty-traffic mode, where each \ac{UAV}-\ac{UE} is assumed to transmit a packet of a certain size. The same network topology as described in \tref{tab:sim_parameter} for the full-buffer mode is still applied herein for the bursty-traffic mode. Differently, \ac{UAV}-\acp{UE} in the bursty-traffic mode are assumed to enter the network according to a Poisson distribution with an arrival rate $\lambda$ as 2.5 \ac{UE}/s/cell. All active \ac{UAV}-\acp{UE} transmit 4\,Mb data and leave the network once the transmission is finished. 
 Additional parameters configured for the bursty-traffic mode are included in \tref{tab:sim_bursty}. Furthermore, in the full-buffer simulation as discussed before, it is clear that we assume the algorithm or the computing center knows the information of the channel gains among all UAV-BS pairs, i.e. $\mathbf G$, so that all the UAV-UEs can be optimized jointly. We denote it as ``centralized application of algorithms'', which may require considerable resources for \ac{UAV}-\acp{UE} to perform channel estimation and feedback the estimated channel information to the computing center. It is also possible to ``de-centralize'' the algorithm, e.g., that the optimizations are done for individual groups of UAV-UEs. For the bursty-traffic mode, we will discuss both centralized and de-centralized applications of algorithms in the sequel.

\textit{1) Centralized application of algorithms:}

\begin{table}[t]
  \centering
  \caption{Additional parameters applied together with \tref{tab:sim_parameter} for the bursty-traffic mode.}
  \scalebox{1}{
  \begin{tabular}{ll}
  \hline
  \hline
  \multicolumn{2}{c}{Main additional parameters for bursty-traffic mode}\\
  \hline
  Bandwidth $B$   & 9\,MHz      \\
  Arrive rate $\lambda$ of \ac{UAV}-\acp{UE} & 2.5\,\ac{UE}/s/cell \\
  Packet size & 4\,Mb \\
  {Updating time interval } & 0.02\,s \\
  Simulation time  & 10\,s\\
  \hline\hline
  \end{tabular}}
  \label{tab:sim_bursty}
  \end{table}

  \begin{figure*}
    \begin{minipage}{\textwidth}
    \begin{center}
      \redfbox{\subfigure[]{\includegraphics[width=0.47\textwidth]{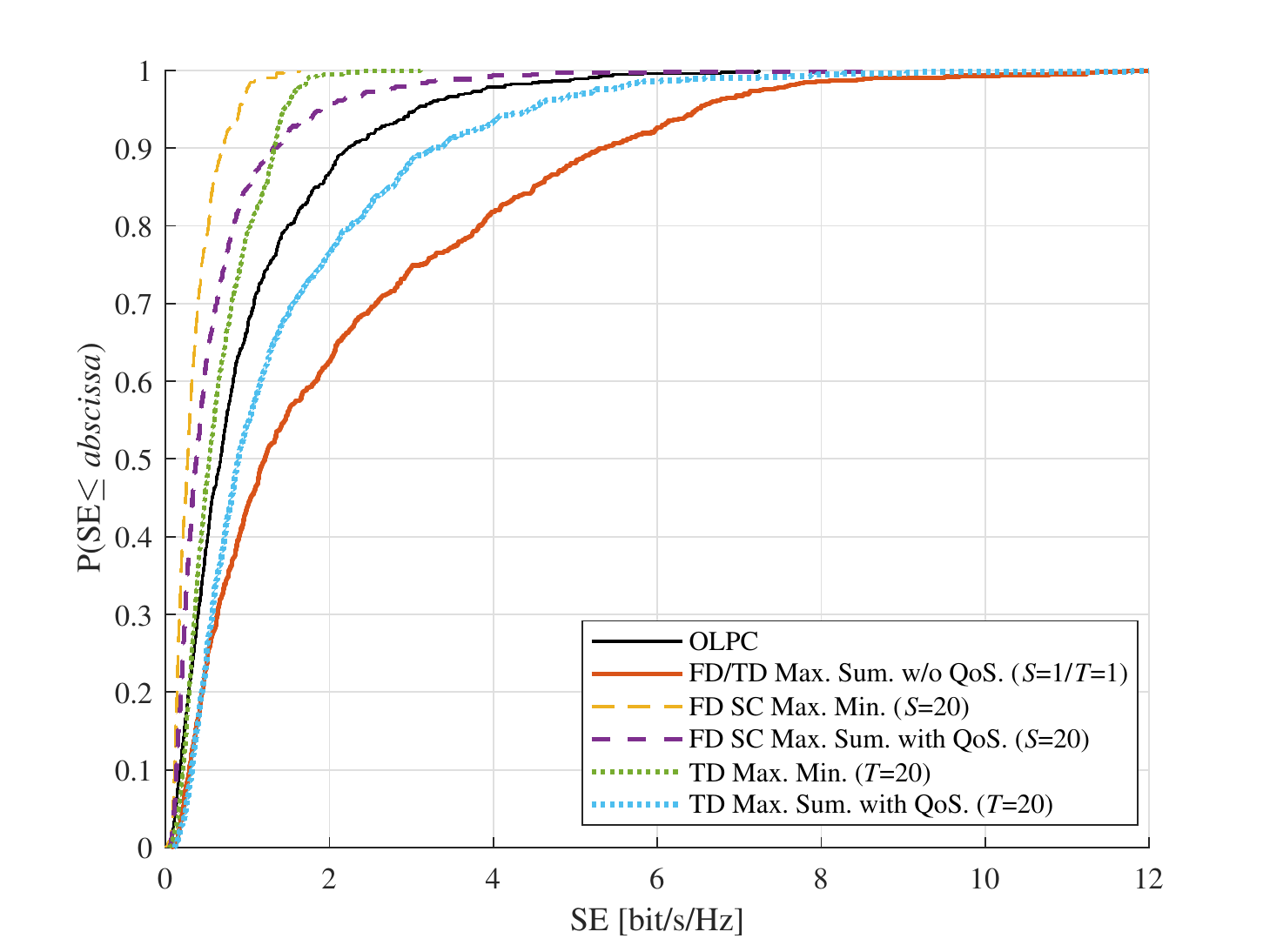}{\psfrag{F(x)}[l][l][0.65]{Omnidirectional}}}}
      \redfbox{\subfigure[]{\includegraphics[width=0.47\textwidth]{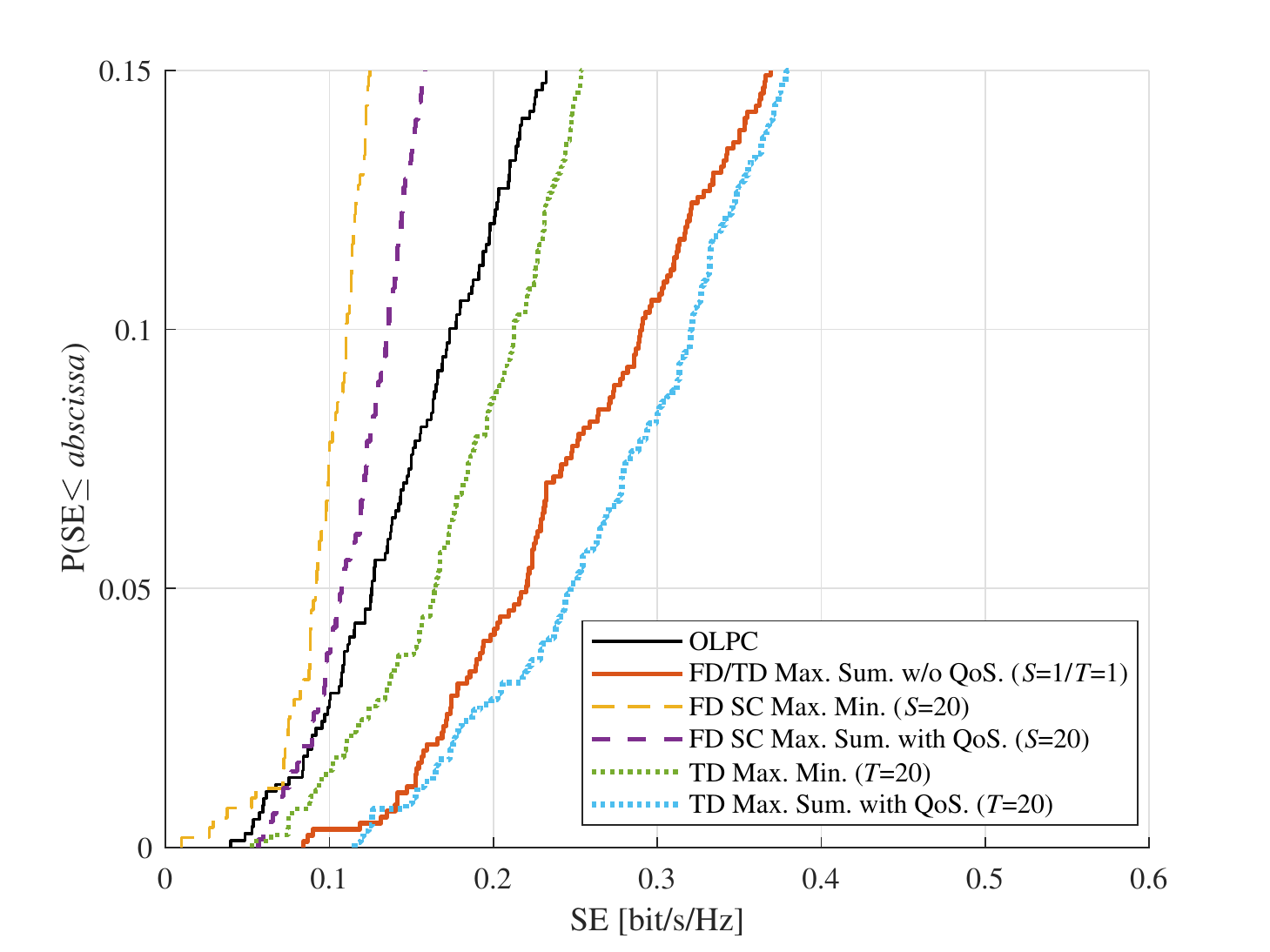}}}
    \end{center} 
    \caption{Performance of different algorithms applied in the bursty-traffic mode. (a) \acp{CDF} of \acp{SE} of all individual \ac{UAV}-\acp{UE}  (b) Zoomed figure. \label{fig:burstytraffic}}
    \begin{center}
    \redfbox{\subfigure[]{\includegraphics[width=0.47\textwidth]{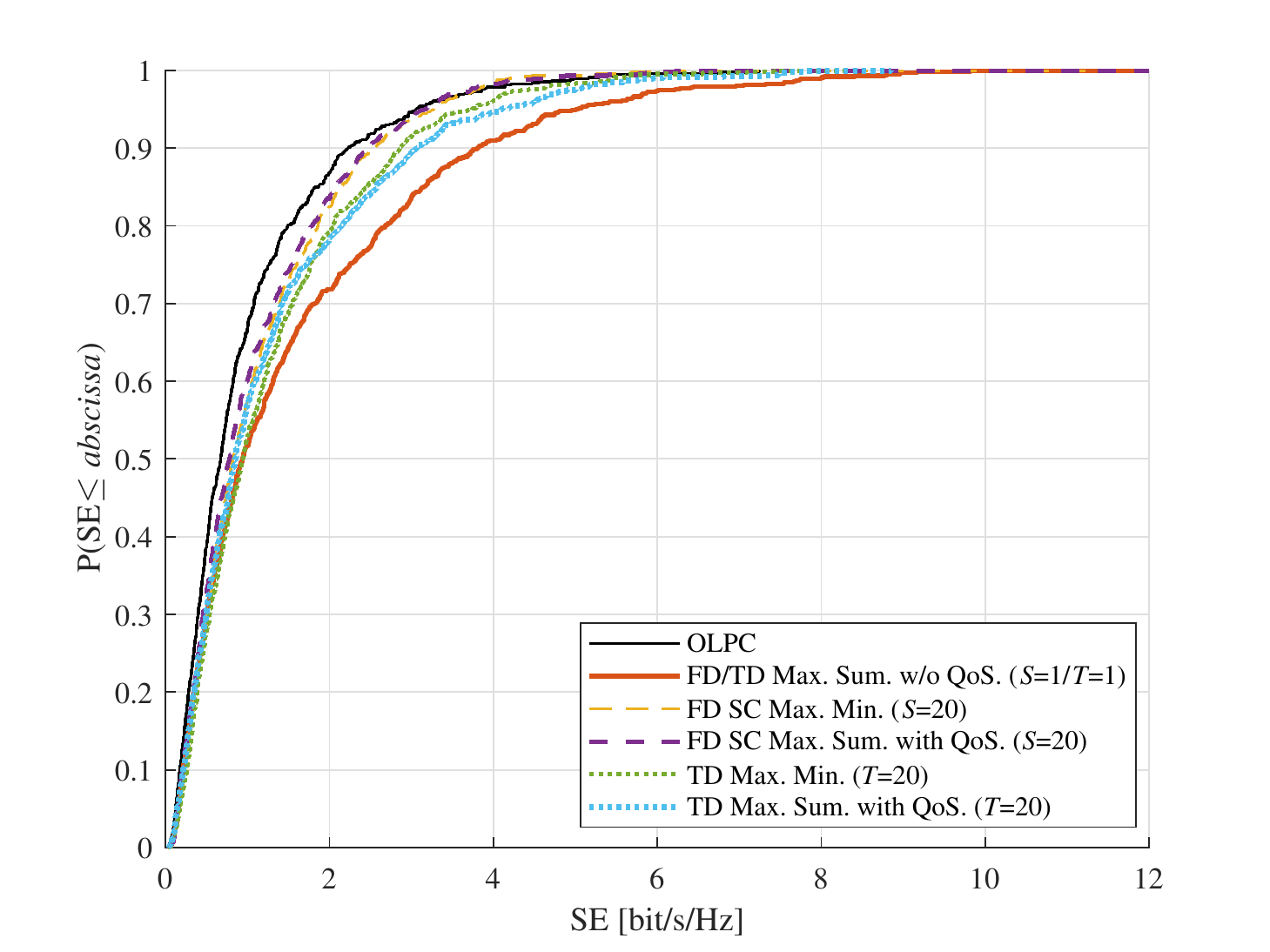}{\psfrag{F(x)}[l][l][0.65]{Omnidirectional}}}}
    \redfbox{\subfigure[]{\includegraphics[width=0.47\textwidth]{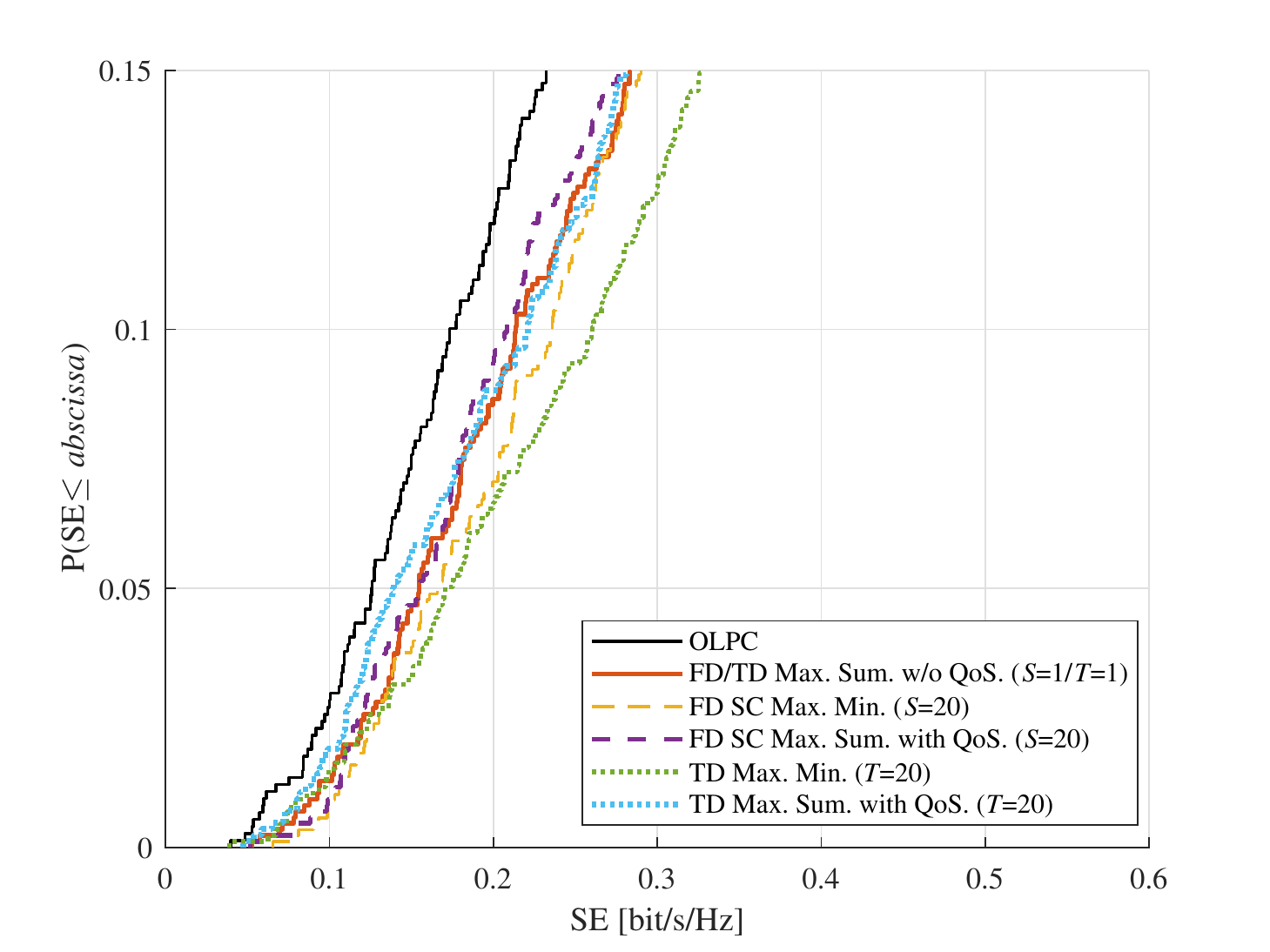}}}
  \end{center} 
  \caption{Performance of different algorithms cluster-wise applied in the bursty-traffic mode. (a) \acp{CDF} of \acp{SE} of all \ac{UAV}-\acp{UE}. (b) Zoomed figure. \label{fig:burstytraffic_clustered}}
\end{minipage}
    \end{figure*}


\fref{fig:burstytraffic} illustrates the \acp{CDF} of \acp{SE} of \ac{UAV}-\acp{UE} for different algorithms applied.\footnote{It can be seen in \tref{tab:full_buffer_mode} that Max-Min ($S$=1/$T$=1) has worse performances for all groups of \ac{UAV}-\acp{UE}. Also considering that the uplink \ac{SC} constraint is practically required, we thus omit Max-Min ($S$=1/$T$=1) and FD Max-Min ($S$=20) in the simulation of bursty-traffic mode. } Specifically, the \ac{SE} $R_\text{bursty}$ of a \ac{UAV}-\ac{UE} in the bursty-traffic mode is calculated as 
  \begin{equation}
    \begin{aligned}
    R_\text{bursty} = \frac{DataSize}{B (t_\text{out} - t_\text{in})}
    \end{aligned}
    \label{eq:burstyrate}
    \end{equation} where $t_\text{in}$ and $t_\text{out}$ are the time instants when the \ac{UE} enters into and leaves the system, respectively, and $DataSize$ [bit] is the total amount of data transmitted by the \ac{UAV} in the period from $t_\text{in}$ to $t_\text{out}$. Moreover, the first 1.6\,seconds of the system is not considered to exclude the ``warming-up'' stage. \fref{fig:burstytraffic} provides the following insights.
    \begin{itemize}
\item  It is interesting to observe from \fref{fig:burstytraffic} that in the bursty-traffic mode FD\,SC\,Max-Min and FD\,SC\,Max-Sum with QoS have smaller performances, i.e. $R_\text{bursty}$ in \eqref{eq:burstyrate}, for all \ac{UAV}-\acp{UE} than that of \ac{OLPC}, although in the full-buffer mode they can achieve larger minimum \acp{SE} compared to that of \ac{OLPC}. {This is because the \acp{SE} of \ac{UAV}-\acp{UE} in better channel conditions are limited to maximize the minimum \ac{SE} or to meet the QoS constraint in each updating time interval. In other words, the mean (or sum) SE of UAV-UEs is decreased in an updating interval, as indicated in \fref{fig:fullbuffer}(b). Consequently, \ac{UAV}-\acp{UE} in relatively good channel conditions stay in the system for a longer time compared to the case, e.g., in the Max-Sum without QoS constraint, where they can transmit at much higher rates. Moreover, since these UAV-UEs in relatively good conditions tend to be scheduled with higher priorities compared to those UAV-UEs in relatively bad channel conditions, the average \acp{SE} of UAV-UEs in bad channel conditions are also decreased due to longer inactive periods, although when they are active they indeed have higher minimum \acp{SE}, finally leading to smaller averaged \acp{SE} of all \ac{UAV}-\acp{UE} in the bursty mode.}
This is similarly true for observing that TD-based algorithms have better performance than that of their corresponding FD algorithms, e.g. from FD\,SC\,Max-Min to TD Max-Min, and for that TD/FD Max-Sum with QoS have better performances than TD/FD Max-Min in the simulation case herein. 
\item However, it is not necessarily that the larger the sum (or mean) \acp{SE} in individual updating time intervals are, the better the overall performance is. For example, comparing the curves of TD Max-Sum with QoS and Max-Sum without QoS, it can be observed that TD Max-Sum with QoS has better performance for \ac{UAV}-\acp{UE} at lower percentiles, although Max-Sum without QoS indeed can achieve largest sum \ac{SE} at each updating time interval. {This is because in each updating time interval of Max-Sum without QoS constant, a certain number of \acp{UE} are almost muted with zero rates, as indicated in \fref{fig:fullbuffer}(c).} This means that a \ac{SE} balance between \ac{UAV}-\acp{UE} in bad and good channel conditions in each updating time interval must be properly tuned to achieve a good system performance in the bursty-traffic mode, especially when the \acp{QoS} or \acp{SE} of \acp{UE} at lower percentiles are key evaluation metrics. 
    \end{itemize}
\tref{tab:burstytraffic} summarizes the performance gains of the algorithms applied in the bursty-traffic mode for different groups of  UAV-UEs compared to that of \ac{OLPC}. It can be seen that TD Max-Sum with QoS is a good algorithm with the best edge \ac{UE} performance and still not bad overall performance. {By tuning the QoS constraints, it is possible to achieve an optimized balance between \ac{UAV}-\acp{UE} in bad and good channel conditions for pre-defined key evaluation metrics.}

\begin{table}[]
  \centering
    \caption{Performance gains (in percentage) for the $H$-th percentile \acp{SE} of all \ac{UAV}-\acp{UE} in the bursty traffic mode compared to \ac{OLPC}. The top three best algorithms are bolded.}
    \label{tab:my-table}
    \redfbox{
    \scalebox{0.75}{
    \begin{tabular}{l|r|r|r|r}
    \hline
   {\diagbox{Algorithms}{$H$}} & {10} & {20} & {50} & {100} \\ \hline
                        %
                        FD/TD Max-Sum w/o QoS. ($S$/$T$=1) &   \textbf{66\%}  &  \textbf{60\%}  &   \textbf{79\%}  &   \textbf{85\%}  \\ \hline 
                        FD SC Max-Min ($S$=20) &   -37\%  &   -50\%  &   -60\%  &   -77\%  \\ \hline 
                        FD SC Max-Sum with QoS. ($S$=20)  &   -22\%  &   -33\%  &   -44\%  &   \textbf{18\%}  \\ \hline 
                        TD Max-Min ($T$=20)   &  \textbf{22\%}  &   \textbf{2\%}  &   \textbf{-21\%}  &   -57\%  \\ \hline 
                        TD Max-Sum with QoS. ($T$=20)   &  \textbf{83\%}  &  \textbf{58\%}  &   \textbf{32\%}  &   \textbf{73\%}  \\ \hline 
    \end{tabular}}
    }
    \label{tab:burstytraffic}
    \end{table}

\textit{2) Decentralized (locally centralized) application of algorithms:}

We also evaluate the performances of the algorithms when applying them separately to clusters of cells in the network. That is, the whole network is divided into several clusters of cells, and the algorithms are applied for each of these cells without considering the existence of other clusters in each updating time interval{\footnote{\addnew{Note that they are different from the so-called distributed algorithms where the master problems are decomposed into several subproblems that are solved locally and synchronized through message-passing \cite{4395184}. The results are different from that achieved in the centralized or distributed algorithms. Although how to realize distributed algorithms is not addressed in this paper, it is an interesting future work since the overhead can be reduced significantly compared to that of the centralized algorithms.}}}.  After optimizing the resource and power allocation of each cluster, the performance of the whole network is evaluated by combining all clusters.  The main reasons to do so include that \textit{i)} The centralized way requires channel state information among all \acp{UE} and all sector cells, which is probably practically challenging and consumes additional resources for signaling; \textit{ii)} Giving too much fairness among all \ac{UAV}-\acp{UE} may finally degrade the overall performance since the \acp{UE} with relatively good channel conditions may be also limited;
\textit{iii)} Although a \ac{UAV}-\ac{UE} may cause interference to many cells, it is still true that the badly affected ones are cells relatively close to this \ac{UAV}-\ac{UE} in terms of distance and beam direction. In the simulation considered herein where individual \ac{UAV}-\acp{UE} are assumed to use the best beam with a 60$^\circ$ of \ac{HPBW}, each of three co-site sectored cells is grouped as a cluster as illustrated in \fref{fig:cells}, i.e., the whole network with 48 sector cells is divided into 16 clusters.

  \fref{fig:burstytraffic_clustered} illustrates the \ac{CDF}s of the \acp{SE} of all \ac{UAV}-\acp{UE} for different algorithms applied cluster-wise. It is interesting to find that all the proposed algorithms have better performances compared to that of \ac{OLPC}. Specially, FD\,SC\,Max-Min, FD\,SC\,Max-Sum with QoS and TD Max-Min are improved significantly compared to that as illustrated in \fref{fig:burstytraffic}. This is because fairness in each updating time interval is decreased which leads to a larger sum \ac{SE} hence improving the final bursty performance. However, the performances of TD Max-Sum with QoS and Max-Sum without QoS decrease slightly compared to their centralized versions as shown in \fref{fig:burstytraffic}. {The reason is that the centralized TD Max-Sum with QoS and Max-Sum without QoS can already achieve relatively good performances for both cell-edge and cell-center \ac{UAV}-\acp{UE}, i.e., the fairness problem is not that severe as in, e.g., FD\,SC\,Max-Min, and decentralized applications of them instead decreases the performance since the UAV-UEs were not jointly optimized.}
  

\subsection{Overall discussions} \label{sect:discuss}
Based on the extensive simulations in both full-buffer mode and bursty-traffic mode conducted for the proposed algorithms, we would like to summarize several points as follows. 

\begin{itemize}
\item The proposed algorithms in the frequency domain and/or time domain can obtain the optimal or suboptimal resource and power allocations for \ac{UAV}-\acp{UE} subject to certain constraints. Generally, time domain algorithms can achieve better performances, since \ac{SC} constraint has to be considered when applying frequency domain algorithms. However, whenever the \ac{SC} constraint is no longer required, frequency domain algorithms are better choices than the time domain algorithms, since higher power densities can be achieved.  
\item In the full-buffer mode which emulates high traffic load, algorithms that maximize the minimum \ac{SE} can have the best minimum \ac{SE}, whereas the \acp{SE} of \ac{UAV}-\acp{UE} in good conditions, e.g. cell-center \ac{UAV}-\acp{UE}, may be limited significantly. Maximizing the sum \ac{SE} without considering QoS constraints can achieve the best sum \ac{SE}. However, the \acp{SE} of some \ac{UAV}-\acp{UE} can be always close to zero. Nevertheless, maximizing the sum \ac{SE} considering QoS constraints is a compromise for both cell-edge and cell-center \ac{UAV}-\acp{UE}. Moreover, it is possible to achieve an optimized value for the target key performance indicator by properly tuning the \ac{QoS} constraint for different service scenarios. 
\item The performance of \ac{UAV}-\acp{UE} in the bursty-traffic mode is different from that in the full-buffer mode. For example, it is not necessarily that the algorithm that maximizes the minimum \ac{SE} in each updating time interval can achieve better averaged \acp{SE} for lower-percentile \ac{UAV}-\acp{UE}.
Overall, performances of all \ac{UAV}-\acp{UE} depend on both the minimum \ac{SE} and sum \ac{SE} achieved in each updating time interval. The algorithm that maximized the sum \ac{SE} in TD with a certain QoS constraint can be considered the best option for bursty-traffic mode transmission. \addnew{It is worth noting that the centralized algorithms need to know all channel gains and require signaling to the central computation unit. The complexity of optimization also increases polynomially with the number of UAV-UEs. To decrease the signaling and computation load, decentralized application of algorithms could be a practical solution.}  
The performance of decentralized frequency-domain algorithms is found to be increased, although they are still lower than that of the centralized TD maximization of sum SE with QoS constraints. Nevertheless, if the time domain algorithm cannot be done, e.g. due to the short updating time interval of the system, decentralized frequency domain algorithms can be good options.

 
\end{itemize}


    







\section{Conclusions}\label{sect5}
In this contribution, different power allocation algorithms have been proposed for the uplink communications of a massive number of cellular-connected \acfp{UAV}. Generally, the time-domain maximization of sum \acf{SE} with a properly tuned \acf{QoS} constraint works satisfactorily for \acp{UAV} in both high and medium/low traffic conditions.
Other algorithms may emphasize different groups of \ac{UAV}-\acp{UE}. Moreover, scheduling multiple \acp{UAV} are favorable for power-limited \acp{UAV} as the maximum power density can be increased. Cluster-wise applications, with lower computation and signaling loads, can also increase the performance of frequency domain algorithms. 
Future work can investigate joint optimization of scheduling and power allocation for multi-cell multi-UAVs. \addnew{Moreover, since UAV engines usually consume a lot of energy, optimizing the flight trajectories of UAVs is also an important aspect to be (jointly) considered.}  


\setlength{\itemsep}{0em}
\renewcommand{\baselinestretch}{1.08}
\patchcmd{\thebibliography}
  {\settowidth}
  {\setlength{\parsep}{0pt}\setlength{\itemsep}{0pt plus 0.1pt}\settowidth}
  {}{}

  \normalem

  \setlength{\itemsep}{0em}
\renewcommand{\baselinestretch}{1}
\bibliographystyle{IEEEtran}
\bibliography{IEEEabrv,reference}

\begin{IEEEbiography}[{\includegraphics[width=1in,height=1.25in,clip,keepaspectratio]{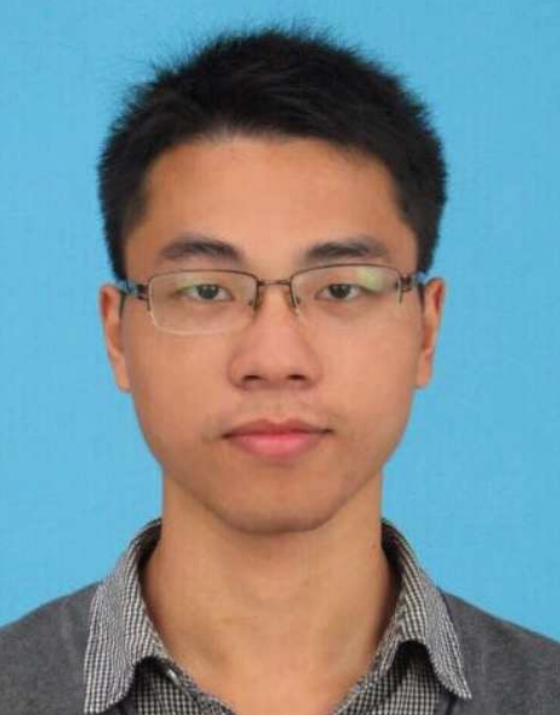}}]{Xuesong Cai} (Senior Member, IEEE) received the B.S. degree and the Ph.D. degree (with distinction) from Tongji University, Shanghai, China, in 2013 and 2018, respectively. In 2015, he conducted a three-month internship with Huawei Technologies, Shanghai, China. He was also a Visiting Scholar with Universidad Polit\'ecnica de Madrid, Madrid, Spain in 2016. From 2018-2022, he conducted several postdoctoral stays at Aalborg University and Nokia Bell Labs, Denmark, and Lund University, Sweden. He is now an Assistant Professor and a Marie Skłodowska-Curie Fellow at Lund University in Communications Engineering, closely cooperating with Ericsson. His research interests include radio propagation, high-resolution parameter estimation, over-the-air testing, resource optimization, and radio-based localization for 5G/B5G wireless systems.

  Dr. Cai was a recipient of the China National Scholarship (the highest honor for Ph.D. Candidates) in 2016, the Outstanding Doctorate Graduate awarded by the Shanghai Municipal Education Commission in 2018, the Marie Skłodowska-Curie Actions (MSCA) ``Seal of Excellence'' in 2019, the EU MSCA Fellowship (ranking top 1.2\%, overall success rate 14\%) and the Starting Grant (success rate 12\%) funded by the Swedish Research Council in 2022. He was also selected by the ``ZTE Blue Sword-Future Leaders Plan'' in 2018 and the ``Huawei Genius Youth Program'' in 2021.
  \end{IEEEbiography}
  \begin{IEEEbiography}[{\includegraphics[width=1in,height=1.25in,clip,keepaspectratio]{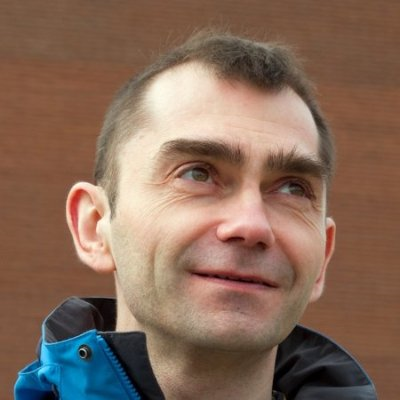}}]{Istv\'an Zsolt Kov\'acs} (Member, IEEE) received his B.Sc. from “Politehnica” Technical University of Timişoara, Romania in 1989, his M.Sc.E.E. from École Nationale Supérieure des Télécommunications de Bretagne,  France in 1996, and his Ph.D.E.E. in Wireless Communications  from Aalborg University, Denmark in 2002. Currently he is senior research engineer at Nokia, Aalborg, Denmark, where he conducts research on machine learning-driven radio resource management and on radio connectivity enhancements for non-terrestrial and aerial vehicle communications, in LTE and 5G networks.
  \end{IEEEbiography}

  \begin{IEEEbiography}[{\includegraphics[width=1in,height=1.25in,clip,keepaspectratio]{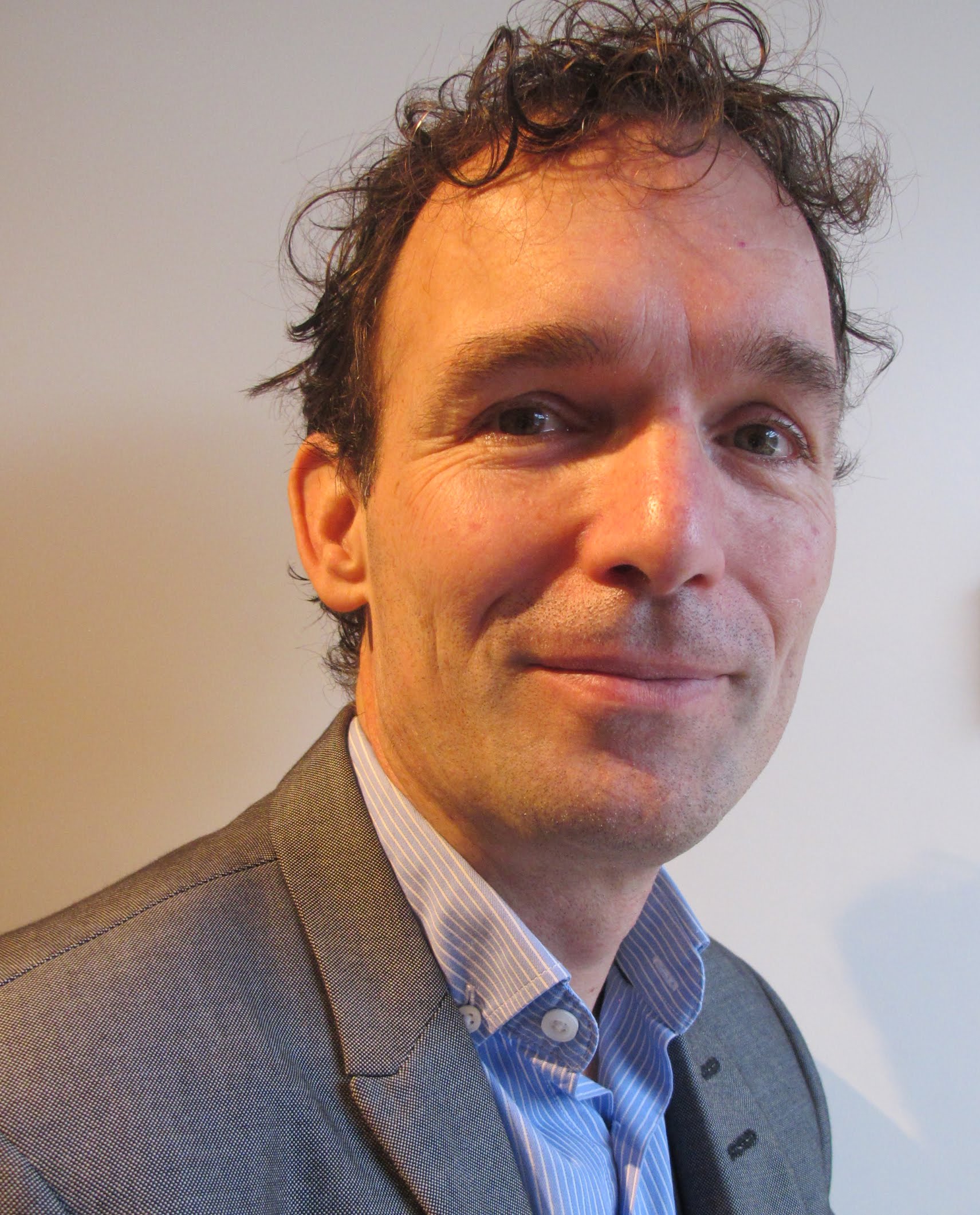}}]{Jeroen Wigard} received the M.Sc. degree in electrical engineering from Technische Universiteit Delft, Netherlands, in 1995, and the Ph.D. degree on the topic of handover algorithms and frequency planning in frequency hopping GSM networks from Aalborg University, Denmark, in 1999. He joined Nokia Aalborg, Denmark, where he worked on radio resource management-related topics for 2G, 3G, 4G, and 5G networks. He is currently with Nokia Aalborg (former Nokia Networks Aalborg) and is involved in studies and 5G standardization related to UAVs and nonterrestrial networks (NTN). He has authored and coauthored over 60 journals and conference papers and holds more than 100 patent applications.
    
  \end{IEEEbiography}

  \begin{IEEEbiography}[]{Rafhael Amorim} 
  \end{IEEEbiography}

  \begin{IEEEbiography}[{\includegraphics[width=1in,height=1.25in,clip,keepaspectratio]{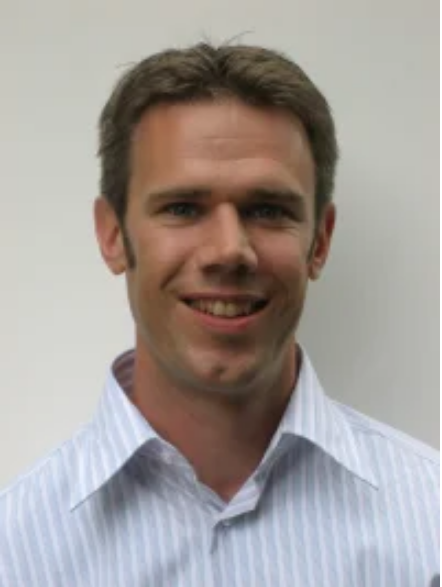}}]{Fredrik Tufvesson} (Fellow, IEEE) 
    received the Ph.D. degree from Lund University, Lund, Sweden, in 2000.
    
    After two years at a startup company, he joined the Department of Electrical and Information Technology, Lund University, where he is currently a Professor of radio systems. He has authored around 100 journal articles and 150 conference papers. His main research interest is the interplay between the radio channel and the rest of the communication system with various applications in 5G/B5G systems, such as massive multiple-input multiple-output (MIMO), mmWave communication, vehicular communication, and radio-based positioning.
    
    Dr. Tufvesson’s research has been awarded the Neal Shepherd Memorial Award for the Best Propagation Paper in the IEEE
    TRANSACTIONS ON VEHICULAR TECHNOLOGY and the IEEE Communications Society Best Tutorial Paper Award.

    \end{IEEEbiography}

  \begin{IEEEbiography}[{\includegraphics[width=1in,height=1.25in,clip,keepaspectratio]{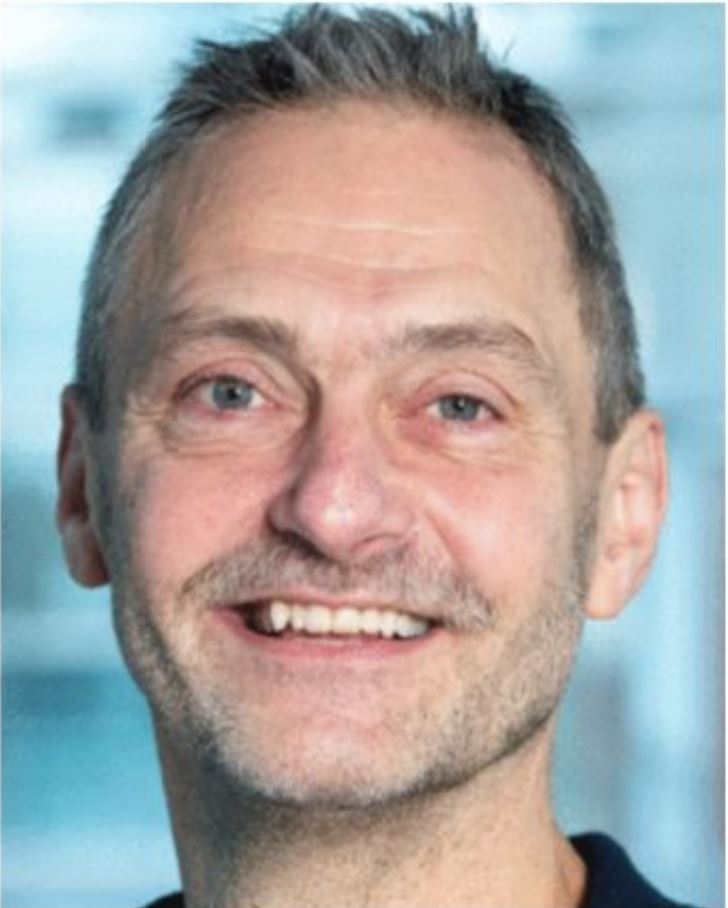}}]{Preben E. Mogensen} received the M.Sc. and Ph.D. degrees from Aalborg University, in 1988 and 1996, respectively. Since 1995, he has been a part-time Associate with Nokia. Since 2000, he has been a Full Professor with Aalborg University, where he is currently leading the Wireless Communication Networks section. He is also a Principal Scientist with Nokia Bell Labs and a Nokia Bell Labs Fellow. He has coauthored over 450 articles in various domains of wireless communication. His Google Scholar H-index is 68. His current research interests include the 5G industrial IoT and technology components towards 6G.
  \end{IEEEbiography}

\end{document}